\documentclass[preprint,superscriptaddress,amsmath,amssymb,aps,prab,longbibliography,nofootinbib,titlepage]{revtex4-1}

\usepackage{bm}
\usepackage{slashed}
\usepackage{graphicx}
\usepackage{natbib}
\usepackage{hyperref}
\usepackage{cleveref}
\usepackage{mathrsfs}
\usepackage{gensymb}
\usepackage[usenames,dvipsnames]{xcolor}
\usepackage{newtxtext,newtxmath}
\usepackage{makecell}

\hypersetup{colorlinks=true,citecolor=blue,urlcolor=blue,filecolor=blue}

\newcommand{\rmd}{\mathrm{d}}
\newcommand{\micron}{{\upmu\mathrm{m}}}

\newcommand{\abs}[1]{\left| #1 \right|}

\newcommand{\avg}[1]{{\left\langle {#1} \right\rangle}}
\newcommand{\zmf}{{\text{zmf}}}

\newcommand{\chirp}{c}
\newcommand{\Prob}{\tsf{P}}

\newcommand{\mcS}{\mathcal{S}}
\newcommand{\mcM}{\mathcal{M}}
\newcommand{\mcI}{\mathcal{I}}

\newcommand{\lcm}{{\scriptscriptstyle -}} 
\newcommand{\lcp}{{\scriptscriptstyle +}}
\newcommand{\lcpm}{{\scriptscriptstyle \pm}}
\newcommand{\lcmp}{{\scriptscriptstyle \mp}}
\newcommand{\lcperp}{{\scriptscriptstyle \perp}}

\newcommand{\newa}{a_{\text{rms}}}
\definecolor{bk1}{rgb}{0.12, 0.56, 1.0}
\newcommand{\bi}{\begin{itemize}}
\newcommand{\ei}{\end{itemize}}

\newcommand{\tsf}[1]{\textsf{#1}}

\newcommand{\trm}[1]{\textrm{#1}}

\newcommand{\vkap}{\kappa}
\newcommand{\eps}{\varepsilon}

\newcommand{\gaugea}{a}

\begin{document}

\title{From local to nonlocal: higher fidelity simulations of photon emission in intense laser pulses}

\author{T. G. Blackburn}
\email{tom.blackburn@physics.gu.se}
\affiliation{Department of Physics, University of Gothenburg, SE-41296 Gothenburg, Sweden}
\author{A. J. MacLeod}
\author{B. King}
\affiliation{Centre for Mathematical Sciences, University of Plymouth, Plymouth, PL4 8AA, United Kingdom}

\date{\today}

\begin{abstract}
State-of-the-art numerical simulations of quantum electrodynamical (QED) processes in strong laser fields rely on a semiclassical combination of classical equations of motion and QED rates, which are calculated in the locally constant field approximation.
However, the latter approximation is unreliable if the amplitude of the fields, $a_0$, is comparable to unity.
Furthermore, it cannot, by definition, capture interference effects that give rise to harmonic structure.
Here we present an alternative numerical approach, which resolves these two issues by combining cycle-averaged equations of motion and QED rates calculated in the locally monochromatic approximation.
We demonstrate that it significantly improves the accuracy of simulations of photon emission across the full range of photon energies and laser intensities, in plane-wave, chirped and focused background fields.
\end{abstract}

\maketitle


\section{Introduction}

The collision of multi-GeV electron beams and intense laser pulses is a promising scenario for precision measurements of quantum electrodynamics (QED) in the strong-field regime, where both the normalised amplitude of the laser, $a_0$, and quantum nonlinearity parameter of the electron, $\chi_e$, exceed unity.
Perturbative QED calculations of the interaction fail once $a_0 \not\ll 1$ and must be replaced by `all-order' approaches, which take the interaction with the strong background field into account exactly~\cite{ritus.jslr.1985,dipiazza.rmp.2012}.
While the theory for this regime is now several decades old~\cite{Nikishov:1964zza}, experiments are limited in number.
In the weakly multiphoton regime, $a_0 \simeq 0.4$, laser-electron collision experiments have observed Compton scattering (photon emission) and trident electron-positron pair creation ~\cite{bula.prl.1996,burke.prl.1997}.
At higher values of $a_{0}$, but small $\chi_{e}$, they have observed photon emission in the classical regime (nonlinear Thomson scattering)~\cite{sarri.prl.2014,sakai15,khrennikov15,yan17} and at $a_{0}\simeq 10$, radiation reaction (multiple photon emission) in the nonlinear classical~\cite{cole.prx.2018} and quantum regimes~\cite{poder.prx.2018}.
However, as yet, there are no experimental measurements charting the transition between the perturbative, multiphoton, and nonlinear regimes, $0.1 \lesssim a_0 \lesssim 10$ at $\chi_e \simeq 1$.
This is likely to change in the near future, as increasing interest in strong-field QED has led to planned experiments that will combine conventional electron accelerators with intense optical lasers~\cite{meuren.exhilp.2019,luxe}.

The transition regime represents a particular challenge for theory and simulation.
A perturbative approach is not sufficient once $a_0 \not\ll 1$.
However, neither is an approach based on the \emph{locally constant field approximation} (LCFA)~\cite{ritus.jslr.1985,erber.rmp.1966}, as this applies only in the opposite limit, $a_0 \gg 1$: this approximation underpins the simulation codes~\cite{ridgers.jcp.2014,gonoskov.pre.2015,lobet.jpc.2016} used to model QED effects in laser-plasma interactions~\cite{elkina.prstab.2011,ridgers.prl.2012,gelfer15,jirka16,stark.prl.2016,grismayer.pre.2017,zhang.pop.2020}, which will be explored in the next generation of multi-petawatt laser facilities~\cite{papadopoulos.hpl.2016,weber.mre.2017,gales.rpp.2018,danson.hplse.2019}.
The versatility of the LCFA comes from its local nature and the neglect of interference effects, i.e. the finite size of the spacetime region over which QED processes take place, which requires both $a_0 \gg 1$ and $a_0^3 / \chi_e \gg 1$; the limitations of doing so have been thoroughly discussed in the literature~\cite{harvey.pra.2015,blackburn.pop.2018,dipiazza.pra.2018,ilderton.pra.2019,king.pra.2020}.
Experiments that aim at precision measurements of strong-field QED demand precision simulations of the interaction.
However, in the transition regime, the error made by simulations based on LCFA rates is unacceptably large.

In this paper, we present a simulation framework that overcomes these issues by using the \emph{locally monochromatic approximation} (LMA) instead.
This achieves greater accuracy by taking into account interference effects at the scale of the laser wavelength, which is possible provided that the laser pulse is relatively unchanged by the collision with a probe electron beam.
To do this, we combine classical trajectories, defined on a cycle-averaged basis, with probability rates that treat the background `locally' as a monochromatic plane wave, with an amplitude and frequency that can vary in space and time.
As such, we exchange the ability of the LCFA to model an \emph{arbitrary} electromagnetic field for significantly increased accuracy in the modelling of \emph{plane-wave-like} fields.
While plane-wave rates have already been used in numerical modelling and analysis~\cite{cain,bamber.prd.1999,hartin.ijmpa.2018,luxe}, their derivation from strong-field QED has only recently been formalised by \citet{heinzl.pra.2020}, who combine a slowly varying envelope approximation~\cite{Narozhnyi:1996qf,McDonald:1997,seipt.pra.2011,seipt.jpp.2016} with a `local' expansion in the interference phase~\cite{Nikishov:1964zza,ritus.jslr.1985,harvey.pra.2015,ilderton.pra.2019,dipiazza.pra.2018,dipiazza.pra.2019}.

The derivation of the LMA in \citet{heinzl.pra.2020} assumes a plane wave, whereas any experimental configuration will employ a focused laser pulse.
This makes it essential to consider beyond-plane-wave field configurations, for which exact theoretical results are limited in number~\cite{heinzl.prl.2017,heinzl.jpa.2017}.
In order to make progress, we consider the case of plane-wave backgrounds that have a nonlinear dependence on phase, or a `chirp', which results in a localisation of both the wave's amplitude and frequency.
By allowing both the amplitude and wavevector to vary in space and time, we gain analytical insight into the case of a focused background, where this would also be the case.
We then describe how the LMA may be implemented in numerical simulations of photon emission and benchmark their predictions against strong-field QED for pulsed plane waves (unchirped and chirped) as well as with focusing pulses.
For the last of these, we must employ an approximate solution to the Dirac equation~\cite{dipiazza.prl.2014,dipiazza.prl.2016,dipiazza.pra.2017}, which, to the best of our knowledge, has not previously been compared to a simulation.
Our results confirm that simulations based on this framework may be used for precision modelling of experiments, with an accuracy of a few percent in the integrated probability (improving on the accuracy of the LCFA by orders of magnitude in the transition regime), and correct reproduction of harmonic structure in the differential spectrum, which has been identified as an aim of future experiments~\cite{luxe}.

In the following, we use a system of units in which the Planck's reduced constant, the speed of light and the vacuum permittivity are all set to unity: $\hbar = c = \epsilon_0 = 1$.
The electron mass is denoted by $m$.
The fine-structure constant $\alpha$ is related to the elementary charge $e$ by $\alpha = e^2/(4\pi)$.

\section{Theory background}
\label{sec:Theory}

We begin with an explanation of how the full QED plane-wave results are calculated, as well as a summary of the main details arising from the analytical calculation underpinning the LMA.
(Many papers have investigated the effect of pulse shape on nonlinear Compton scattering, see e.g.  \cite{boca09,PhysRevA.81.022125,PhysRevA.83.022101,mackenroth11,PhysRevA.85.062102,PhysRevA.89.032125}.)
For concreteness, we specify from the outset that we will be assuming a background that is a circularly polarised, chirped, plane-wave pulse with potential $A$.
We define the dimensionless potential $\gaugea=eA/m$,
    \begin{align}
    \gaugea(\phi) = a_{0} f\!\left(\frac{\phi}{\Phi}\right) \left[\eps\,\cos b(\phi) + \beta\,\sin b(\phi)\right]
        ,
    \label{eqn:ChirpedBG}
    \end{align}
where $a_{0}$ is the dimensionless intensity parameter~\cite{heinzl.oc.2009} (also called the ``classical nonlinearity'', normalised amplitude or the strength parameter) and $\varepsilon$, $\beta$ are orthonormal polarisation vectors obeying $\varepsilon \cdot \varepsilon = \beta\cdot \beta=-1$.
Throughout, we use lightfront coordinates $x^{\mu} = (x^{\lcp},x^{\lcm},\vec{x}^{\lcperp})^{\mu}$, where $x^{\lcpm} = x^{0} \pm x^{3}$, $\vec{x}^{\lcperp} = (x^{1},x^{2})$, $x^{\lcpm} = 2 x_{\lcmp}$ and $\vec{x}^{\lcperp} = - \vec{x}_{\lcperp}$.
The function $f(\phi/\Phi)$ is the pulse envelope which depends on the lightfront phase $\phi = \vkap \cdot x$ (where $\vkap_{\mu} = \delta^{+}_{\mu} \vkap_{+}$ is the background wavevector), and the pulse phase duration, $\Phi$, is related to the number of cycles, $N$, via $\Phi = 2 N$.
The function $b(\phi)$ describes the chirp of the background.
For a pulse without chirp, $b$ is linear in $\phi$, i.e. $b''(\phi)=0$ for all $\phi$.
(In the following, we will pick $b(\phi)=\phi$ for the unchirped case.)

We use the scattering matrix approach~\cite{Schwartz:2013pla} to calculate the probability of single nonlinear Compton scattering from a single incoming electron colliding with a plane-wave background.
We can write the scattering matrix element as:
    \begin{equation}
    \tsf{S}_{r',r;l} = -ie
        \int\,
        \rmd^{4}x ~\overline{\Psi}_{p',r'}(x) \slashed{\epsilon}^{*}_{k,l} e^{i k \cdot x} \Psi_{p,r}(x),
    \label{eqn:Sfi1}
    \end{equation}
where $\slashed{\epsilon}^{*}_{k,l}$ is the polarisation of the emitted photon with 4-momentum $k$ and $\Psi_{p,r}~(\overline{\Psi}_{p',r'})$ is the Volkov wavefunction~\cite{Volkov:1935zz} of the incoming (outgoing) electron:
    \begin{align}
    \Psi_{p,r}(x) &= \left(1 + \frac{m \slashed{\vkap}\slashed{\gaugea}}{2\,\vkap\cdot p}\right)u_{p,r}\,\mbox{e}^{iS_{p}(x)},
    &
    S_p(x) &= p \cdot x + \int^{\phi} \!\rmd y\, \frac{2 m p \cdot \gaugea(y) - m^2 \gaugea^2(y)}{2 \vkap\cdot p}.
    \label{eq:Volkov}
    \end{align}
The matrix element can be simplified to:
    \begin{multline}
    \tsf{S}_{r',r;l} =
        \tilde{C}
        \int_{\phi_{i}}^{\phi_{f}} \!\rmd\phi \,
        \bar{u}_{r'}
        \left[
            \Delta \slashed{\epsilon}^{\ast}_{k,l} + \frac{m}{2\,\vkap\cdot p}\left(\frac{ \slashed{\gaugea}\slashed{\vkap}\slashed{\epsilon}^{\ast}_{k,l}}{1-s}+\slashed{\epsilon}^{\ast}_{k,l}\slashed{\gaugea}\slashed{\vkap}\right)
        \right]
        u_{r}
        \exp\!\left[
            \frac{i}{\eta_{0}(1-s)}
            \int^{\phi}_{\phi_i} \!\rmd y \, \frac{k\cdot \pi(y)}{m^{2}}
        \right]
    \label{eqn:MatrixElement}
    \end{multline}
where $s=\vkap\cdot k/\vkap\cdot p$ is the lightfront momentum fraction of the emitted photon, $\eta_{0} = \kappa \cdot p/m^{2}$ is the initial energy parameter of the probe electron, $\tilde{C}$ contains normalisation constants, the instantaneous electron momentum is given by
    \begin{equation}
    \pi(y) = p - m\gaugea(y) + \vkap \frac{2\,m p\cdot \gaugea(y) - m^2 \gaugea^{2}(y)}{2\,\vkap\cdot p},
    \label{eq:ClassicalKineticMomentum}
    \end{equation}
and the regularising factor $\Delta = 1 - k\cdot \pi/k\cdot p$ incorporates all the contributions from phases outside of the integral.
The total probability can be written:
    \begin{equation}
    \Prob = \frac{\alpha}{\eta_{0}} \frac{1}{2^{4}\pi^2}
        \int\!
        \rmd^2 \vec{r}^\perp \rmd s \,
        \frac{s}{1-s}
        \langle|\tsf{S}_{r',r;l}|^{2}\rangle_{\tsf{pol.}},
    \label{eqn:P1}
    \end{equation}
where $\vec{r}^\perp = \vec{k}^\perp/(m s) - \vec{p}^\perp/m$ contains the shifted perpendicular momentum.
Here ``$^{\perp}$'' indicates directions perpendicular to the background propagation direction and $\avg{\cdot}_{\tsf{pol.}}$ indicates an average over initial and sum over final polarisation states.
The numerical results in exact QED are calculated by evaluating \cref{eqn:P1} directly: the matrix element in \cref{eqn:Sfi1} was evaluated using photon polarisation eigenstates of the background \cite{baier75a} and spin states in the Lepage-Brodsky convention~\cite{brodsky.pr.1998}.

Rather than direct numerical evaluation, some of the integrals in \cref{eqn:P1} can be evaluated analytically  by generalising the locally monochromatic approximation~\cite{heinzl.pra.2020} to arbitrarily chirped plane-wave pulses.
In the following, we present an overview of this approach, and direct the reader to \cref{app:LMA} for details. 

The background field is given by \cref{eqn:ChirpedBG}. 
For the LMA to approximate the emission spectrum well, the envelope function $f(\phi/\Phi)$ should be \emph{slowly varying} with respect to the carrier frequency, implying that $\Phi^{-1} \ll \min\!\left[b'(\phi)\right]$ (i.e. $\Phi \gg 1$ for the unchirped case, which corresponds to a many-cycle pulse).
However, in this work, we also include the chirp.
Therefore we will also make a ``slowly varying chirp'' approximation (see e.g. \citet{seipt.pra.2015}).
These approximations then allow the squared Kibble mass, $\mu$, which occurs in an exponent, to be integrated over. The Kibble mass takes the form $\mu = 1 + \avg{\vec{a}^2}_\theta - \avg{\vec{a}}^2_\theta$, where $\avg{f}_{\theta}=\theta^{-1}\int^{\phi+\theta/2}_{\phi-\theta/2}f$ denotes a phase-window average.
In the case of a circularly polarised background, the slowly varying (envelope) and rapid (carrier) timescales occur in $\avg{\vec{a}}_\theta$. We can demonstrate the approximation by considering a single component of $\vec{a}$, e.g. $\vec{\eps}\cdot \vec{a}$.
    \begin{equation}
    \avg{ \vec{\eps} \cdot \vec{a}}_{\theta} =
        \frac{a_{0}}{\theta}
        \int^{\phi+\theta/2}_{\phi-\theta/2}
        \! \rmd\phi
        \, f\!\left(\frac{\phi}{\Phi}\right)
        \cos b(\phi).
    \end{equation}
Now, one can introduce a \emph{local frequency scale}, $\omega(\varphi) = b^{\prime}(\varphi)$ and integrate by parts as in \cref{eqn:IntegralsTransform}. The fast timescale of the cosine term is included \emph{exactly}. The remaining terms for the envelope and chirp variations have a size, relative to the leading term, of the order of
    \begin{align}
    &\sim \frac{1}{\Phi} \frac{f'(\phi / \Phi)}{f(\phi / \Phi)},
    &
    &\sim \frac{\omega'(\phi)}{\omega(\phi)},
    \label{eq:LmaError}
    \end{align}
respectively (neglecting a rapidly varying term that appears $\sim\cot b(\phi)$). As long as the magnitudes of both of these are much less than unity, we should expect the slowly varying approximation to be good. (The same arguments apply to the $\vec{\beta}\cdot \vec{\gaugea}$ term, whereas $\avg{\vec{a}^2}_\theta$ is not affected by chirp in a circularly polarised background.) Beyond the additional constraints on the chirp, no further modifications to \cite{heinzl.pra.2020} are required in the derivation (more details are given in Appendix \cref{app:LMA}).

Finally, we arrive at $\Prob^\text{LMA} = \int\! d\tau \,W^\text{LMA}$, where:
    \begin{equation}
    W^\text{LMA} =
        \sum_{n=1}^{\infty}
        \int_{0}^{s_{n,*}(\tau)}
        \! \rmd s \,
        \frac{\rmd^{2}\Prob^\text{mono}_{n}\left[\newa(\tau),\eta(\tau)\right]}{\rmd\tau\,\rmd s}
    \label{eqn:LMARate}
    \end{equation}
where $\newa^{2}(\tau) = q^{2}/m^{2}-1$ and $\eta(\tau)=\omega[\phi(\tau)] \eta_{0}$, with $\eta_{0}=\vkap\cdot p/m^2$ the \emph{unchirped} energy parameter.
Here $q=\avg{\pi}$ is the \emph{quasimomentum}, the laser-cycle-average of the instantaneous electron momentum given in \cref{eq:ClassicalKineticMomentum}.
The appearance of a \emph{local} wavevector in $\eta(\tau)$ also follows from considering components of the field-strength tensor, $F^{\mu\nu}$, for the chirped pulse in \cref{eqn:ChirpedBG}, which contain terms $\sim \kappa^{\mu}(\phi)\, \partial a^{\nu}/ \partial b$, where $\kappa^{\mu}(\phi) = b'(\phi)\kappa^{\mu}$.
$\Prob^\text{mono}_{n}$ is the probability of nonlinear Compton scattering into the $n$th harmonic in a monochromatic background, $\tau$ is the proper time, related to the phase by $d\tau/d\phi=1/(m \eta_{0})$.
The approximation is \emph{locally} monochromatic because the intensity and energy parameter occurring in the monochromatic probability now take the (cycle-averaged) \emph{local} value at the position of the electron. The integrand is given explicitly by \cref{eqn:NLCProbFinal} for nonlinear Compton scattering. Unlike the monochromatic case, here the harmonic range is phase-dependent:
    \begin{align}
    s_{n,\ast}(\tau) &= \frac{s_{n}(\tau)}{1+s_{n}(\tau)},
    &
    s_{n}(\tau) &= \frac{2n \eta(\tau)}{1+\newa^{2}(\tau)}, \label{eqn:sndef}
    \end{align}
where $s_{n}(\tau)$ is the edge of the classical (nonlinear) harmonic range.

To obtain the probability of Compton scattering in a focused laser background, we must use some approximation, as analytical solutions to the Dirac equation in a realistic focused laser background are unavailable (some progress has recently been made in this direction: see e.g.~\cite{heinzl.prl.2017,heinzl.jpa.2017}). One method is to find an approximate solution to the Dirac equation using a WKB expansion in a small parameter $\gamma^{-1}$, where $\gamma$ is the initial relativistic gamma factor of the incident electron~\cite{dipiazza.prl.2014,dipiazza.prl.2016,dipiazza.pra.2017}. Then assuming $\gamma \gg a_{0}$, for a head-on collision of the electron probe with the focused laser pulse, one can write:
    \begin{equation}
    \Prob^\text{2D} = \int\! \rmd^2 \vec{x}^\perp
        \, \rho(\vec{x}^\perp) \Prob[\newa(\vec{x}^\perp), \eta(\vec{x}^\perp)],
    \label{eq:Prob2d}
    \end{equation}
where $\rho$ is the electron probe areal density and the plane-wave probability, $\Prob$ from \cref{eqn:P1}, now has an intensity parameter which can depend on the perpendicular spatial co-ordinate. 
\vspace{0.5cm}

\section{Implementation in numerical simulations}
\label{sec:Simulations}

    \begin{figure*}
    \centering
    \includegraphics[width=0.8\linewidth]{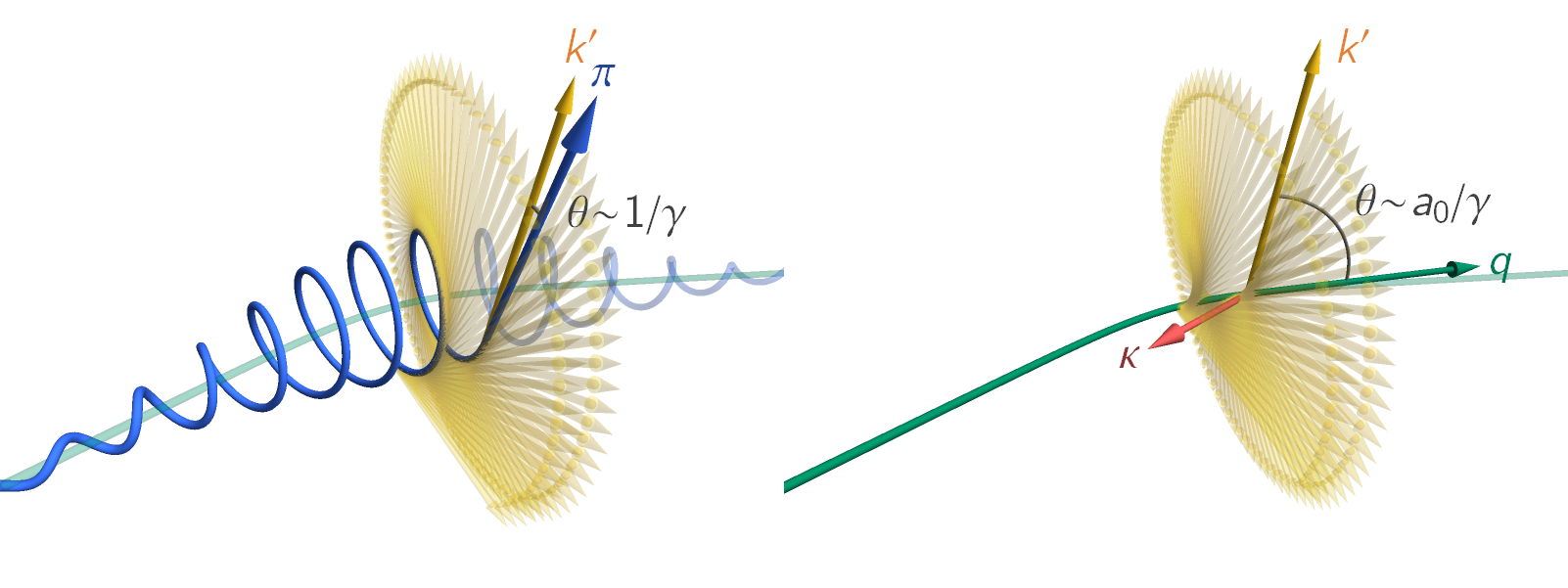}
    \caption{%
        Illustration of two ways to model photon emission by an electron interacting with a high-intensity laser.
        In the locally constant field approximation (left), the kinetic momentum $\pi^\mu$ of the electron (blue) plays the essential role, appearing in the equation of motion, the conservation of momentum, and the emission rate, the latter via the quantum parameter $\chi$.
        In the locally monochromatic approximation (right), it is the \emph{quasi-momentum} $q \equiv \avg{\pi}$ (green) that appears in the conservation of momentum and the emission rate, via the parameters $\newa = \sqrt{q^2/m^2 - 1}$ and $\eta = \kappa \cdot q / m^2$.
        The yellow arrow denotes the emitted photon, momentum $k$, and the red arrow the wavevector of the laser background $\kappa$.
        }
    \label{fig:Concept}
    \end{figure*}

The inclusion of strong-field QED processes in numerical simulations, such as the particle-in-cell~\cite{ridgers.jcp.2014,gonoskov.pre.2015} or particle-tracking codes~\cite{cain,guinea.pig,bamber.prd.1999} used in plasma and beam physics, is based on a semiclassical treatment of particle dynamics, which combines classical trajectories with the use of probability rates~\cite{blackburn.rmpp.2020}.
This is motivated by the appearance of the \emph{classical} kinetic momentum $\pi$, \cref{eq:ClassicalKineticMomentum}, in the QED scattering probability, via the exponent of the Volkov wavefunction, \cref{eq:Volkov}.
(This occurs because the Volkov solution is identical to the semiclassical solution of the Dirac equation in a plane-wave background.)
This permits the probability, \cref{eqn:P1}, to be approximated as the integral $\Prob \simeq \int\! W \,\rmd\tau$, where $W \geq 0$ is interpreted as a probability rate, which can depend, inter alia, on the instantaneous momentum and field amplitude.

The approximations applied to the probability rate affect what dynamical quantities must be obtained from the classical trajectory.
In the locally constant field approximation, for example, the rate $W = W[\chi(\tau)]$, where the quantum nonlinearity parameter $\chi(\tau) = e \abs{F_{\mu\nu}[x(\tau)] \pi^\nu(\tau)} / m^3$ ~\cite{ritus.jslr.1985}.
Furthermore, the conservation of momentum for the scattering may be written such that it constrains the kinetic, rather than asymptotic, momenta.
Thus the classical trajectory must be defined in terms of kinetic momentum $\pi$, i.e. instantaneously, and obtained from the Lorentz force equation $\rmd \pi_\mu/ \rmd \tau = -e F_{\mu\nu} \pi^\nu / m$ and $\rmd x^\mu / \rmd \tau = \pi^{\mu} / m$.
This is illustrated on the left-hand side of \cref{fig:Concept}: the classical trajectory is well-defined at all timescales, including that of the laser carrier wave.
The angular structure of the photon emission arises from two sources: the oscillation of the trajectory ($\theta \simeq a_{0} / \gamma$ for $\gamma \gg a_0 \gg 1$) and the intrinsic beaming of the emission around the instantaneous velocity, the latter being of characteristic size $\theta \sim 1 / \gamma$~\cite{baier,blackburn.pra.2020}.
The former is the dominant contributor in the regime $a_0 \gg 1$, which is consequently where the LCFA is expected to be valid.

The rate in the locally monochromatic approximation, by contrast, is derived assuming that the envelope of the potential, rather than the potential itself, is slowly varying.
Averaging over the fast timescale, the laser period, means that the quantity that enters the rate, and also the conservation of momentum, is not the kinetic momentum directly, but rather the \emph{quasimomentum} $q \equiv \avg{\pi}$~\cite{ritus.jslr.1985,harvey.pra.2009}.
In a plane wave, $\pi = p - m\gaugea + \kappa (2 m p \cdot \gaugea - m^2\gaugea^2) / (2 \kappa \cdot p)$ and $\pi^2 = m^2$, whereas $q = p + \kappa m^2 \newa^2 / (2 \kappa \cdot p)$ and $q^2 = m^2 (1 + \newa^2)$, for $\newa^2 \equiv - \avg{\gaugea^2}$.
In contrast to the LCFA case, the rate is a function of two parameters: the normalised amplitude (or intensity parameter), $\newa$, and the energy parameter $\eta \equiv \kappa \cdot p / m^2$, both locally defined.
(The root-mean-square quantum parameter follows as $\chi_\text{rms} = \newa \eta$.)
Both may be obtained from $q$ as follows: $\newa = \sqrt{(q/m)^2 - 1}$ and $\eta = \kappa \cdot q / m^2$.
An equation of motion for the quasimomentum may be obtained by separating the Lorentz force equation (in a focused, pulsed electromagnetic wave) into quickly and slowly varying components and isolating the latter.
The result is the relativistic ponderomotive force equation~\cite{quesnel.pre.1998}:
    \begin{equation}
    \frac{\rmd \vec{q}}{\rmd t} = -\frac{m^2}{2 q^0} \frac{\partial \newa^2}{\partial \vec{r}}
    \label{eq:PonderomotiveForce}
    \end{equation}
where $q^0 = [m^2(1 + \newa^2) + \abs{\vec{q}}^2]^{1/2}$.
The slowly varying components of the position are determined by
    \begin{equation}
    \frac{\rmd \vec{r}}{\rmd t} = \frac{\vec{q}}{q^0}.
    \label{eq:PositionEvolution}
    \end{equation}
The trajectory obtained from these two equations does not include the fast oscillation at the timescale of the laser period, as shown on the right-hand side of \cref{fig:Concept}.
This does not mean that the physical effect of that oscillation is lost: it is accounted for in the emission rate.
To see this more clearly, note that at fixed $s$, in the limit $a_{0} \gg 1$, there is a most probable harmonic index $n = \newa^2 s / [\eta (1 - s)]$~\cite{seipt.prl.2017}.
Combining this relation with the conservation of quasimomentum, which reads $k_\perp^2/m^2 = 2 n \eta s (1-s) - s^2 (1 + \newa^2)$ for $p_\perp = 0$, one finds that the most probable emission angle is $\theta \simeq \newa / \gamma$ for $\gamma \gg a_0 \gg 1$~\cite{seipt.prl.2017} (see also~\cite{harvey.pra.2009}).
Thus an equivalent angular structure emerges, provided that the classical trajectory is parametrised in terms of quasimomentum.
The conceptual differences between LCFA- and LMA-based simulations are summarized in \cref{tbl:Differences}.

    \begin{table}
    \centering
    \begin{ruledtabular}
    \begin{tabular}{ccc}
        & LCFA & LMA \\
        rate derived for & constant, crossed field & \makecell{monochromatic plane wave \vspace{-0.5em}\\ (\emph{fast quiver motion here})}  \vspace{0.5em}\\
        and controlled by & \makecell{instantaneous momentum $\pi^\mu$ \vspace{-0.5em}\\via quantum parameter $\chi_e$} & \makecell{quasimomentum $q^\mu = \avg{\pi^\mu}$ \vspace{-0.5em}\\ via $\newa$ and $\eta$} \vspace{0.5em}\\
        equation of motion &  \makecell{Lorentz force:\\ $\dfrac{\rmd \pi_\mu}{\rmd \tau} = -\dfrac{e F_{\mu\nu} \pi^\nu}{m}$ \\ (\emph{fast quiver motion here})} & \makecell{ponderomotive force:\\$\dfrac{\rmd \vec{q}}{\rmd t} = -\dfrac{m^2}{2 q^0} \dfrac{\partial \newa^2}{\partial \vec{r}}$}
    \end{tabular}
    \end{ruledtabular}
    \caption{Overview of the conceptual differences between LCFA- and LMA-based simulations of photon emission in strong laser pulses.}
    \label{tbl:Differences}
    \end{table}

The emission of photons, and its effect on this trajectory, is modelled in the following way.
At any particular timestep, we have the electron quasimomentum $q$ and position $r$ from the classical equations of motion, as well as the local values of the laser normalised amplitude $a_\text{rms}(r)$, wavevector $\kappa(r)$ and polarisation (taken to be circular throughout).
In fact, $\kappa$ and $q$ are sufficient to determine the properties of the emission, as they define the two invariant parameters, $a_\text{rms}$ and $\eta$, that control the rate and the conservation of momentum.
This is given by
    \begin{equation}
    q + n \kappa = q' + k,
    \label{eq:MomentumConservation}
    \end{equation}
where $q'$ is the electron quasimomentum after the scattering, $k$ is the momentum of the emitted photon, and $n$ is the harmonic index (the net number of laser photons absorbed).
The emission rates themselves control $n$ and subsequently $s \equiv \kappa \cdot k / \kappa \cdot q$, the lightfront momentum fraction.
Given $n$, $s$ and $q$, it is a matter of kinematics to determine $k$ and then $q'$.
Our Monte Carlo algorithm is as follows:
(i) advance the electron trajectory by solving \cref{eq:PonderomotiveForce,eq:PositionEvolution},
(ii) evaluate, at every timestep, the probability of emission and pseudorandomly decide whether to emit a photon or not,
and on those timesteps where emission takes place,
(iii) select a harmonic index $n$ with probability $W_n / W$, where $W_n$ is the partial rate and $W = \sum_{n = 1}^\infty W_n$ is the total rate,
(iv) sample $s$ from the partial spectrum $(\rmd W_n / \rmd s) / W_n$,
(v) determine $k$ given $n$, $s$ and $q$
and (vi) reset the electron quasimomentum from $q$ to $q'$.

The probability that emission takes place in small interval of lab time $\Delta t$ is given by $\Prob = W \Delta \tau$ and $\Delta \tau = \Delta t (m / q^0)$ is the equivalent interval of proper time.
We obtain $W$ by integrating, and then summing, the partial, differential rates of emission $W_{n}$, which are given by~\cite{heinzl.pra.2020}
    \begin{equation}
    \frac{\rmd W_n}{\rmd s} =
        -\alpha m
        \left\{
            J_n^2(z) +
            \frac{\newa^2}{2}
            \left[ 1 + \frac{s^2}{2(1-s)} \right]
            \left[ 2 J_n^2(z) - J_{n-1}^2(z) - J_{n+1}^2(z) \right]
        \right\}.
    \label{eq:EmissionRate}
    \end{equation}
The argument $z$ of the Bessel functions $J_{n}$ (of the first kind~\cite{olver97}) and auxiliary variables are
    \begin{align}
    z^2 &= \frac{4 n^2 \newa^2}{1 + \newa^2}
        \frac{s}{s_n (1-s)}
        \left[ 1 - \frac{s}{s_n (1-s)} \right],
    &
    s_n &= \frac{2 n \eta}{1 + \newa^2}
    \end{align}
and the bounds on $s$ are $0 < s < s_n / (1 + s_n)$.
Note that $s_n$ depends on $a_\text{rms}$ and $\eta$ and is therefore a function of proper time $\tau$, as shown explicitly in \cref{eqn:sndef}.
While the summation should run from $n = 1$ to infinity, it is sufficient to sum up to a largest value $n_\text{max} = 10 (1 + \newa^3)$.
In principle, the integration and summation can be done at every timestep, given the particular values of $\newa$ and $\eta$.
However, it is significantly faster to obtain $W$ by interpolating from a lookup table, where $W(\newa, \eta)$ is precalculated over the domain $a_\text{rms}^\text{min} < \newa < a_\text{rms}^\text{max}$ and $\eta_\text{min} < \eta < \eta_\text{max}$.
The upper bounds are fixed by the problem space under consideration; we have taken $a_\text{rms}^\text{max} = 10$ and $\eta_\text{max} = 2$ in our code. The lower bounds are chosen such that alternative sampling strategies may be used.

First, if $\newa < a_\text{rms}^\text{min} \ll 1$, only the first harmonic, $n = 1$, contributes significantly to the probability.
In this limit, the rate may be obtained analytically:
    \begin{align}
    W &\simeq W_1 + O(\newa^4),
    &
    W_1 &=
        \frac{\alpha m \newa^2}{2 \eta}
        \left[
            \frac{2 + 8\eta + 9\eta^2 + \eta^3}{(1 + 2\eta)^2}
            -\frac{2 + 2\eta - \eta^2}{2 \eta} \ln(1 + 2\eta)
        \right].
    \label{eq:LowARate}
    \end{align}
Second, if $\eta < \eta_\text{min} \ll 1$, we may take the classical limit, whereupon the partial rates become:
    \begin{align}
    \begin{split}
    \frac{\rmd W_n}{\rmd v} &\simeq
        \frac{\alpha m n \eta}{1 + \newa^2}
        [\newa^2 J_{n-1}^2(z) + \newa^2 J_{n+1}^2(z) - 2(1+\newa^2) J_n^2(z)]
        + O(\eta^2),
    \\
    z^2 &= \frac{4 \newa^2 n^2 v (1-v)}{1 + \newa^2},
    \end{split}
    \label{eq:LowEtaRate}
    \end{align}
but where we fix $v = s (1 + s_n) / s_n$ to be $0 < v < 1$.
\Cref{eq:LowEtaRate}, integrated over $0 < v < 1$ and summed over $n = 1$ to $n_\text{max}$, is tabulated over the same range $a_\text{rms}^\text{min} < \newa < a_\text{rms}^\text{max}$.
In our implementation, $a_\text{rms}^\text{min} = 0.02$ and $\eta_\text{min} = 10^{-3}$.
Thus at every timestep, the emission probability $\Prob = W \Delta\tau$ is obtained by interpolating from the appropriate lookup table, or using the limiting analytical expression.
Emission is deemed to occur if a pseudorandom number $R$, drawn from the uniform distribution $U(0,1)$, satisfies $R < \Prob$.

If emission takes place, the next step is to determine $n$ and $s$. 
The former is obtained by solving for $n$, $R' = \sum_{i=1}^n W_i / W$, where $R'$ is another pseudorandom number drawn on the unit interval $U(0,1)$.
In our implementation, the total rate of emission $W$ is already available at this point; however, the sequence of partial rates must be evaluated explicitly, by integrating \cref{eq:EmissionRate} over $s$.
We do this, rather than store a lookup table in $n$ (as well as in $\newa$ and $\eta$), because unlike the total rate, which is needed at every timestep, the partial rates are only needed on emission, which occurs at infrequent intervals.
Once $n$ is fixed, the lightfront momentum fraction transferred, $s$, is obtained by rejection sampling of \cref{eq:EmissionRate}.

The kinematical calculation of $k$ is performed in the zero momentum frame (ZMF), which moves with four-velocity $u = (q + n \kappa) / [m \sqrt{1 + \newa^2 + 2 n \eta}]$ with respect to the lab frame.
In the ZMF, the emitted photon has momentum $\abs{\vec{k}'_\zmf} = m n \eta / \sqrt{1 + \newa^2 + 2 n \eta}$ and polar scattering angle $\cos \theta_\zmf = 1 - s(1 + \newa^2 + 2 n \eta) / (n \eta)$.
The azimuthal angle $\varphi_\zmf$, which is arbitrary for circularly polarised backgrounds, is pseudorandomly determined in $0 \leq \varphi_\zmf < 2\pi$.
Once $\vec{k}_\zmf$ is determined, it may be boosted back to the lab frame, where $q'$ follows from \cref{eq:MomentumConservation}.

\section{Benchmarking}

While LMA rates have already been implemented in simulation codes used to study laser-electron interactions~\cite{cain,bamber.prd.1999,hartin.ijmpa.2018}, the accuracy of these simulations has not been thoroughly benchmarked against the underlying theory.
Placing quantitative bounds on the error made, is essential for experiments that aim for precision characterisation of strong-field QED processes~\cite{luxe}.
These analyses have been performed for LCFA-based simulations, however: see \cite{harvey.pra.2015,dinu.prl.2016,blackburn.pop.2018} and proposed improvements in \cite{dipiazza.pra.2018,ilderton.pra.2019,king.pra.2020}.
In this section, we compare the results of simulations based on the LMA, as outlined in \cref{sec:Simulations}, with QED theory calculations without the LMA, for photon emission in a pulsed, plane-wave background.
We focus on the transition regime $a_0 \sim 1$, where currently existing approaches based on the LCFA are likely to fail.
The laser pulses we consider are circularly polarised with a cosine-squared temporal envelope: the potential $\vec{\gaugea}(\phi) = a_0 f(\phi) [\vec{x} \cos b(\phi) + \vec{y} \sin b(\phi)]$, where $f(\phi) = \cos^2[\phi/(2N)]$ for $\abs{\phi} < \pi N$.
Here $N$ is the number of cycles corresponding to the total duration of the pulse.
One may estimate the (intensity) full-width-at-half-maximum duration of this pulse as 
$T [\text{fs}] \simeq N \lambda[\micron] / 0.8$.
The function $b(\phi)$ controls the frequency chirping of the pulse and is initially set to $b(\phi) = \phi$ (i.e., unchirped) for the results in \cref{sec:Amplitude}.
The electrons counterpropagate head-on to the laser pulse, with initial energy parameter $\eta_{0} = 0.1$.
This is equivalent to an initial Lorentz factor of $\gamma_0 = 1.638 \times 10^4$ for a laser wavelength of $0.8~\micron$.

The theoretical calculations described in \cref{sec:Theory} are for \emph{single} emission only.
However, for sufficiently large $a_0$ or pulse length $N$, it is possible for the total probability of emission $\Prob$ to exceed unity.
This indicates that higher order processes, including the emission of multiple photons by a single electron, become important.
Simulations model multiple emissions as the incoherent combination of single-vertex processes, transporting the electron classically between emission events.
This is motivated by theoretical calculations of higher order processes which show that part of the probability can be factorised into a product over polarised, first-order processes~\cite{king.prd.2013,dinu.prd.2018,mackenroth.prd.2018}.
Neglecting other contributions, where the intermediate state does not propagate, is expected to be a good approximation if $a_0^2 \, \Delta\phi \gg 1$~\cite{torgrimsson.arxiv.2020}, where $\Delta\phi = 2\pi N$ is the phase duration of the pulse, which allows simulations to model cascades of photon emission and pair creation~\cite{blackburn.rmpp.2020}.
In the present case, we consider only the comparison for single photon emission results.
Therefore, the probability obtained theoretically is interpreted as the average number of emitted photons~\cite{dipiazza.prl.2010}.
As our simulations allow for an arbitrary number of emission events per electron, we obtain equivalent results by artificially disabling recoil, i.e. the electron momentum is not changed self-consistently when a photon is emitted.
The number of emitted photons therefore scales exactly linearly with pulse duration.
This does not apply to the theoretical results.

The symmetries of a plane wave suggest that the photon spectrum is best characterised in terms of the lightfront momentum fraction, $s$, and normalised perpendicular momentum $r_\perp = k_\perp / (m s)$.
These provide proxies for the emitted photon energy $\omega'$ and polar scattering angle $\theta$, respectively: $s = \omega' (1 + \cos\theta) / p^- \simeq \omega' / (m \gamma_0)$ and $r_\perp = (p^- / m) \tan(\theta/2) \simeq \gamma_0 \theta$, where $p^- = m^2 \eta_{0} / \omega_0$ is the initial lightfront momentum of the electron and $\gamma_0$ its Lorentz factor.

\subsection{Pulsed plane waves}
\label{sec:Amplitude}

    \begin{figure*}
    \centering
    \includegraphics[width=\linewidth]{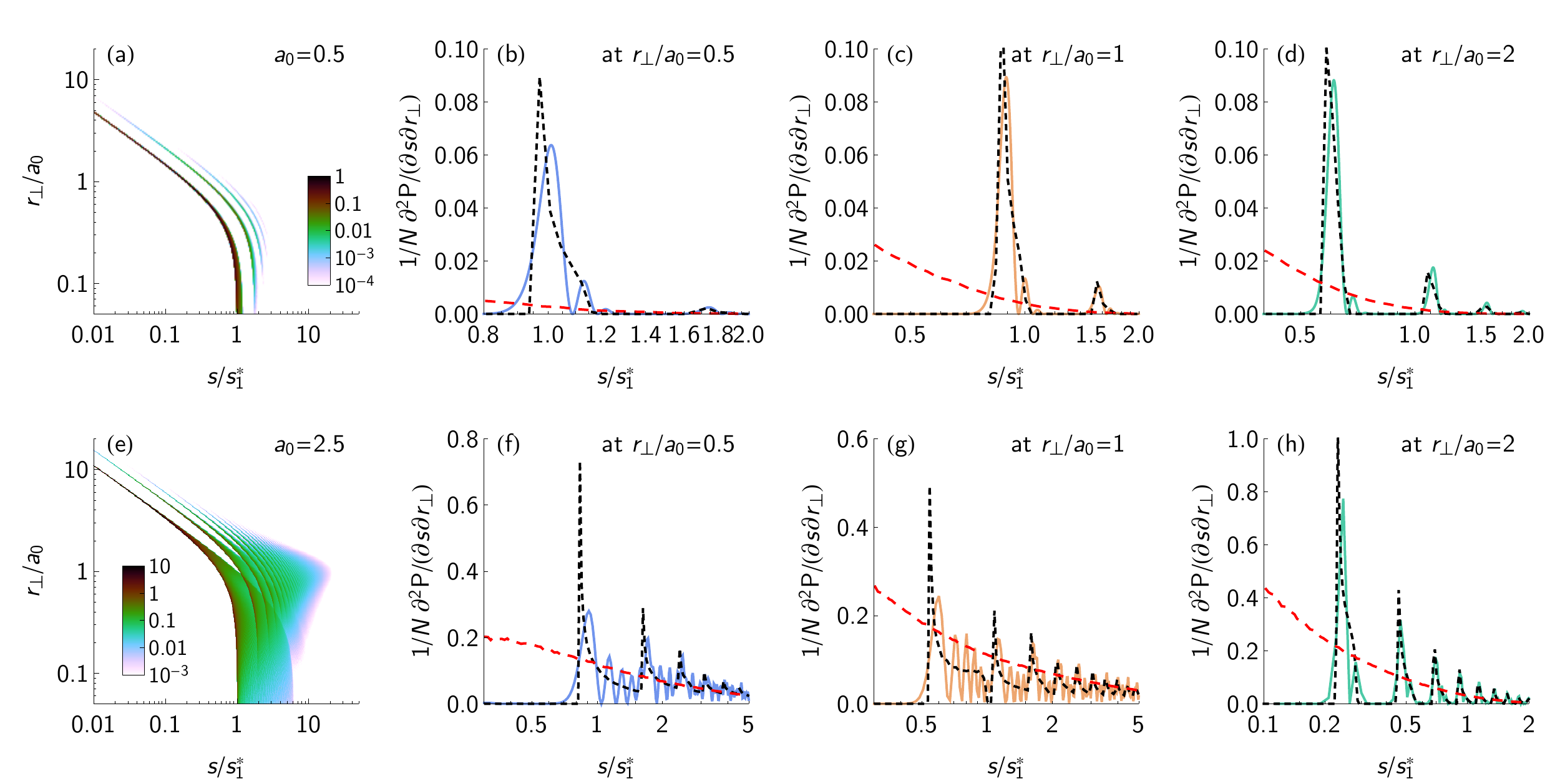}
    \caption{%
        Comparison between theory and simulation results for the double-differential photon spectrum, in the linear regime $a_0 = 0.5$ (upper row) and nonlinear regime $a_0 = 2.5$ (lower row):
        (a) and (e) spectra $\partial^2 \Prob / (\partial s \partial r_\perp)$ from simulations with LMA emission rates (colour scale);
        (b-d) and (f-h) lineouts through the spectrum at fixed $r_\perp$, from theory (solid, coloured) and simulations with LMA (black, dashed) and LCFA (red, dashed) emission rates.
        Here $s^\ast_1 = 2 \eta_{0} / (1 + a_0^2 + 2 \eta_{0})$, which corresponds to the first nonlinear Compton edge, the electron energy parameter $\eta_{0} = 0.1$, and the pulse duration $N = 16$.
        }
    \label{fig:DoubleDifferential}
    \end{figure*}

\Cref{fig:DoubleDifferential}(a) and (e) show photon spectra, double-differential in $s$ and $r_\perp$, obtained from simulations in the linear and nonlinear regimes ($a_0 = 0.5$ and $2.5$ respectively) for a pulse that is $N = 16$ cycles in duration.
In the former case, radiation emission is dominated by the first harmonic, which displays the expected, characteristic energy-angle correlation.
In the latter case, the radiation is composed of a broad range of high harmonics, extending the spectrum to much larger $s$.
The effect of the pulse envelope is evident in the broadening of the first harmonic for small $r_\perp$: recall that the position of the first Compton edge, $s^{\ast}_1 = 2 \eta / (1 + \newa^2 + 2 \eta)$, is phase-dependent through $\newa$ and $\eta$.
We also see that the higher harmonics are predominantly emitted at $r_\perp \simeq a_0$, as expected in the nonlinear regime, whereas for $a_0 = 0.5$, the characteristic $r_\perp < a_0$.

The three plots accompanying each double-differential spectrum compare lineouts at fixed $r_\perp$ against theoretical results.
The simulations capture the position and overall shape of the harmonics well, but miss the subharmonic substructure visible in \cref{fig:DoubleDifferential}(f) and (g) in particular.
This structure arises from interference effects at the scale of the pulse envelope, whereas the LMA accounts only for interference effects at the scale of the wavelength.
The LCFA, by contrast, captures neither, which causes the spectra to be smeared between the clear peaks seen in both the theory and LMA simulation results~\cite{harvey.pra.2015}.

    \begin{figure*}
    \centering
    \includegraphics[width=\linewidth]{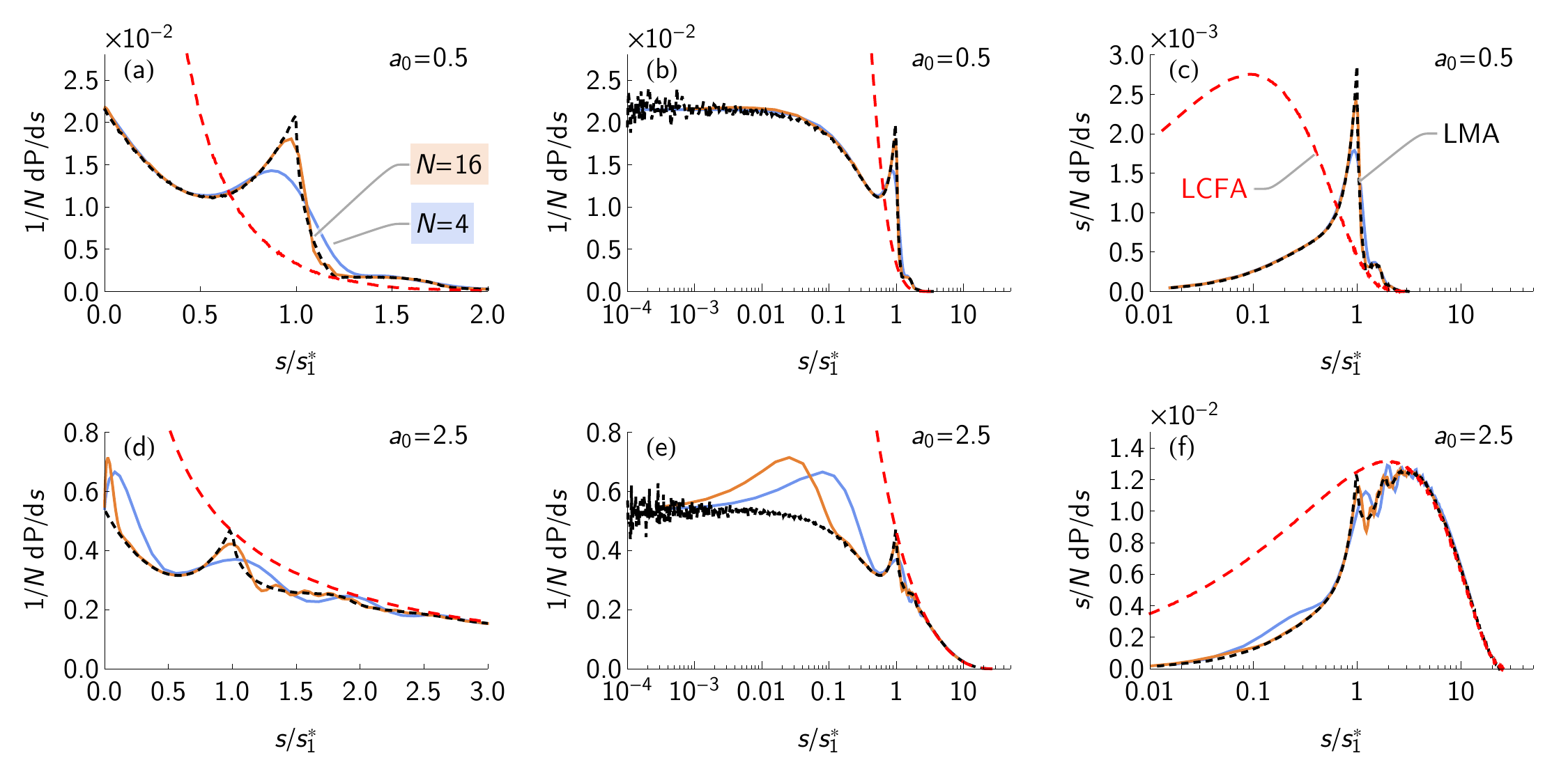}
    \caption{%
        Single differential photon spectra, in the linear regime $a_0 = 0.5$ (upper row) and nonlinear regime $a_0 = 2.5$ (lower row):
        results from QED for a pulse with duration equivalent to $N = 4$ (blue) and $16$ (orange) cycles; and simulations using LMA (black, dashed) and LCFA (red, dashed) emission rates.
        As the spectra are normalised by the duration, and recoil is disabled, the simulation results are independent of $N$ (see text for details).
        Here $s^\ast_1 = 2 \eta_{0} / (1 + a_0^2 + 2 \eta_{0})$, which corresponds to the first nonlinear Compton edge,
        and the electron energy parameter $\eta_{0} = 0.1$.
        }
    \label{fig:SingleDifferential}
    \end{figure*}

Single-differential spectra, i.e. the results from \cref{fig:DoubleDifferential} integrated over $r_\perp$, are shown in \cref{fig:SingleDifferential}.
We compare the simulation results with QED for normalised amplitudes $a_0 = 0.5$ and $2.5$ and for pulse durations equivalent to $N = 4$ and $16$ cycles.
The agreement is much better for the longer pulse, which we expect because the LMA neglects terms of order $1/N$ (see \cref{eq:LmaError} and \cite{heinzl.pra.2020}).
The LMA simulations capture the harmonic structure and correctly reproduce the small-$s$ behaviour of the theory, where the spectrum tends to a constant value $\propto a_0^2 \int\! f^2(\phi) \,\rmd\phi$~\cite{dipiazza.pra.2018,heinzl.pra.2020}.
The LCFA simulations are significantly wrong in this region $s < s^\ast_1$, where we see the characteristic divergence $\propto s^{-2/3}$~\cite{ritus.jslr.1985}.

The intermediate structure, which appears below the first Compton edge for $a_0 = 2.5$, shown in \cref{fig:SingleDifferential}(e), is ponderomotive in origin:
it is radiation from the slow decrease and increase of the electron momentum caused by gradients in the intensity profile~\cite{king.prd.2021}.
While this is accounted for at the level of the classical trajectory in the simulations, its contribution to the emission spectrum is neglected.
The peak moves towards smaller $s$ as $N$ increases and it is eventually lost in the monochromatic limit~\cite{heinzl.pra.2020}.
Integrating over the $s$-weighted probability, shown in \cref{fig:SingleDifferential}(c) and (e), yields the total lightfront momentum transfer from electron to photon.
If $a_0 > 1$, this is dominated by contributions from $s > s^\ast_1$, where the LCFA works well~\cite{blackburn.pop.2018}.
However, it is evident from \cref{fig:SingleDifferential}(c) that the LCFA fails globally for $a_0 < 1$.

    \begin{figure*}
    \centering
    \includegraphics[width=0.8\linewidth]{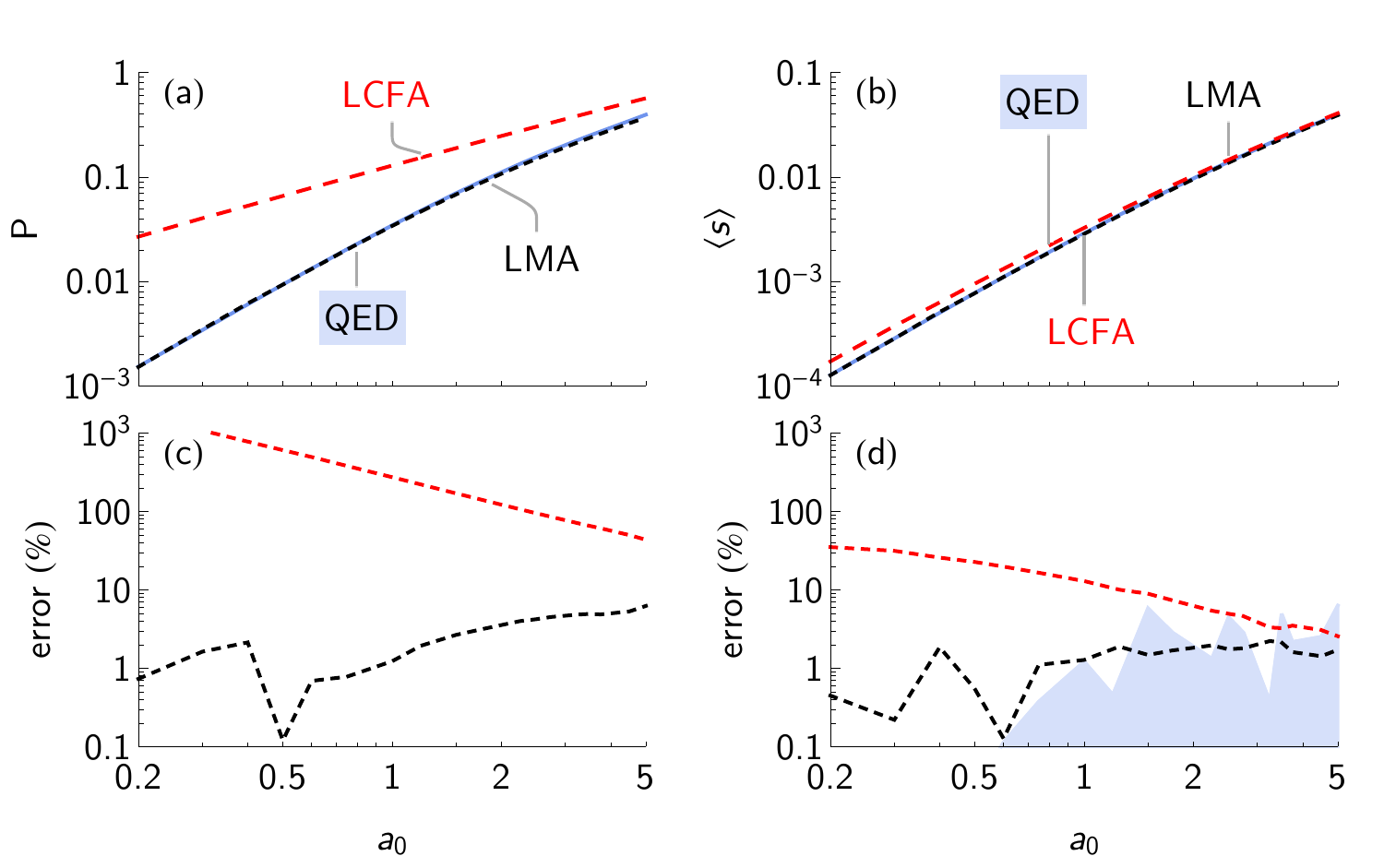}
    \caption{%
        (a) Photon emission probability and (b) average lightfront momentum fraction from QED (blue, solid) and from simulations using LMA (black, dashed) and LCFA (red, dashed) rates.
        Here the pulse duration is equivalent to $N = 4$ cycles and the electron energy parameter $\eta_{0} = 0.1$.
        (c, d) The percentage error of the simulation results, as compared to QED. The blue shaded region gives the estimated accuracy of the QED calculation.
        }
    \label{fig:Probability}
    \end{figure*}
    
Finally, we consider the total probability that a photon is emitted, $\Prob$, and the average lightfront momentum fraction of that photon, $\avg{s} \equiv \int\! s \frac{\rmd \Prob}{\rmd s} \,\rmd s$, as a function of $a_0$ for a four-cycle pulse.
The values obtained from theory and from LMA and LCFA simulations are shown in \cref{fig:Probability}, along with the percentage error made by the simulations.
The LMA-based simulations are accurate at the level of a few per cent across the full range of $a_0$ explored.
The improvement over the LCFA is particularly dramatic for the probability, where the error made is larger than $10\%$ even when $a_0 = 5$.
The average lightfront momentum fraction is more sensitive to the contribution of higher harmonics, i.e. large $s$; as this is where the LCFA works rather well, the accuracy for $\avg{s}$ is better than that for $\Prob$.
However, the LMA simulations are significantly more accurate when $a_0 \lesssim 1$.

\subsection{Chirped pulses}
\label{sec:Chirp}

In \citet{heinzl.pra.2020}, the LMA is derived for a pulse in which the amplitude is slowly varying.
However, a monochromatic plane wave is defined by both an amplitude and a frequency.
By extending the LMA to the situation where both may vary with phase, it becomes possible to simulate radiation generation in chirped laser pulses in the transition regime $a_0 \sim 1$.
In this section we benchmark our simulation results against theory for this case.

    \begin{figure*}
    \centering
    \includegraphics[width=0.8\linewidth]{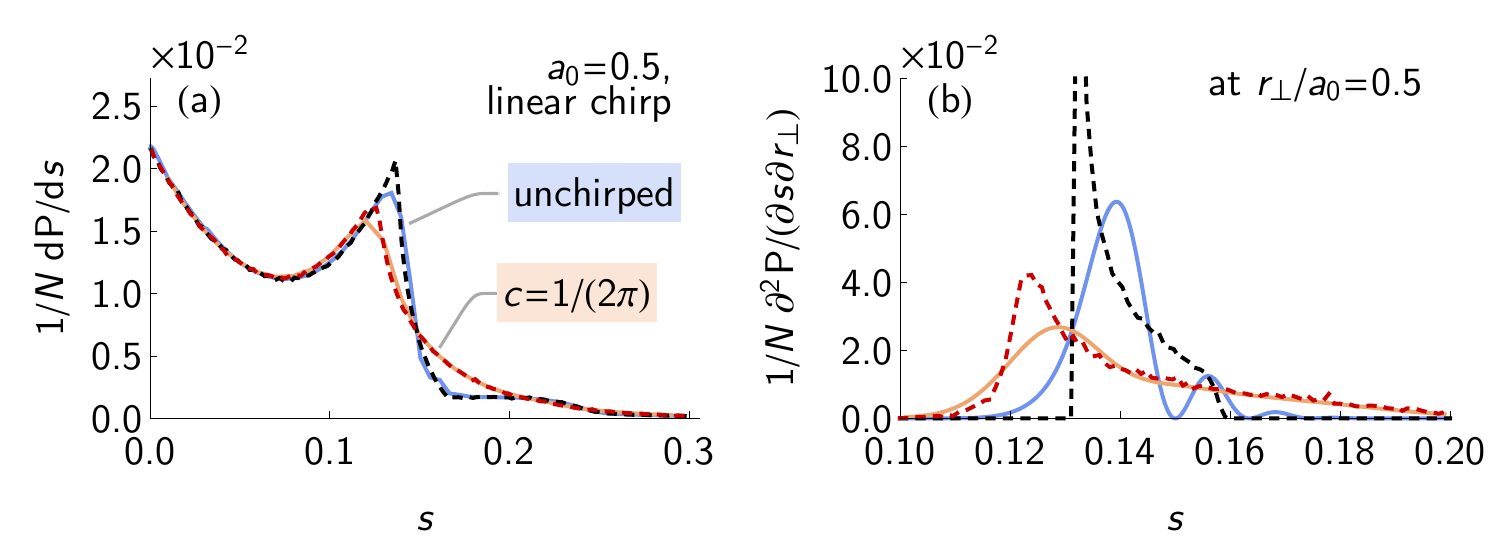}
    \caption{%
        Comparison between simulation (dashed) and QED (solid) results for a linearly chirped pulse with $a_0 = 0.5$ and $N = 16$ (red/orange) and the equivalent unchirped pulse (blue/black).
        The electron energy parameter $\eta_{0} = 0.1$.
        }
    \label{fig:LinearChirp}
    \end{figure*}

The first example we consider is that of a linearly chirped laser pulse, which has potential $\vec{\gaugea}(\phi) = a_0 f(\phi) [\vec{x} \cos b(\phi) + \vec{y} \sin b(\phi)]$, where $f(\phi) = \cos^2[\phi/(2N)]$ for $\abs{\phi} < \pi N$ and $b(\phi) = \phi [1 + \chirp \phi / (2 N) ]$.
The instantaneous frequency, $\omega(\phi) = \omega_0 (1 + \chirp \phi / N)$ for chirp parameter $c$, must be positive throughout the pulse, which imposes the restriction $\chirp < 1/\pi$.
This is consistent with the condition for the chirp to be slowly varying, \cref{eq:LmaError}, which may be cast as $\chirp \ll N / (1 + \pi N)$.
Furthermore, for a particular pulse duration, there is an upper bound on the largest chirp that can be obtained~\cite{silvestro.jqe.1984}.
In our notation, this maximum is given by $\chirp_\text{max} \simeq 10 / N$.
We note that chirping a pulse, which is accomplished by introducing a frequency-dependent phase shift, also changes its duration and peak amplitude; we neglect these such that the only difference between the chirped and unchirped case is the variation of the instantaneous frequency.

We compare the photon spectra obtained from theory and LMA-based simulations for $a_0 = 0.5$, $N = 16$ and $c = 1/(2\pi)$ in \cref{fig:LinearChirp}.
The unchirped results, $c = 0$, are also shown for reference.
The theoretical results are obtained numerically, using \cref{eqn:P1} and the explicit form of the potential $\vec{\gaugea}(\phi)$.
For this case, the electron trajectory can be written in a closed form in terms of Fresnel functions.
In the simulations, a chirp is included by promoting the frequency of the background $\kappa^\mu$ to be a function of phase $\kappa^\mu(\phi)$.
We find that the simulations capture the softening of the harmonic structure evident in the theory results for the chirped pulse.
Lineouts through the theoretical double-differential spectrum at fixed $r_\perp$ demonstrate that chirping smooths out the subharmonic structure; as a consequence, simulation results appear to be more accurate than in the unchirped case.

    \begin{figure*}
    \centering
    \includegraphics[width=0.8\linewidth]{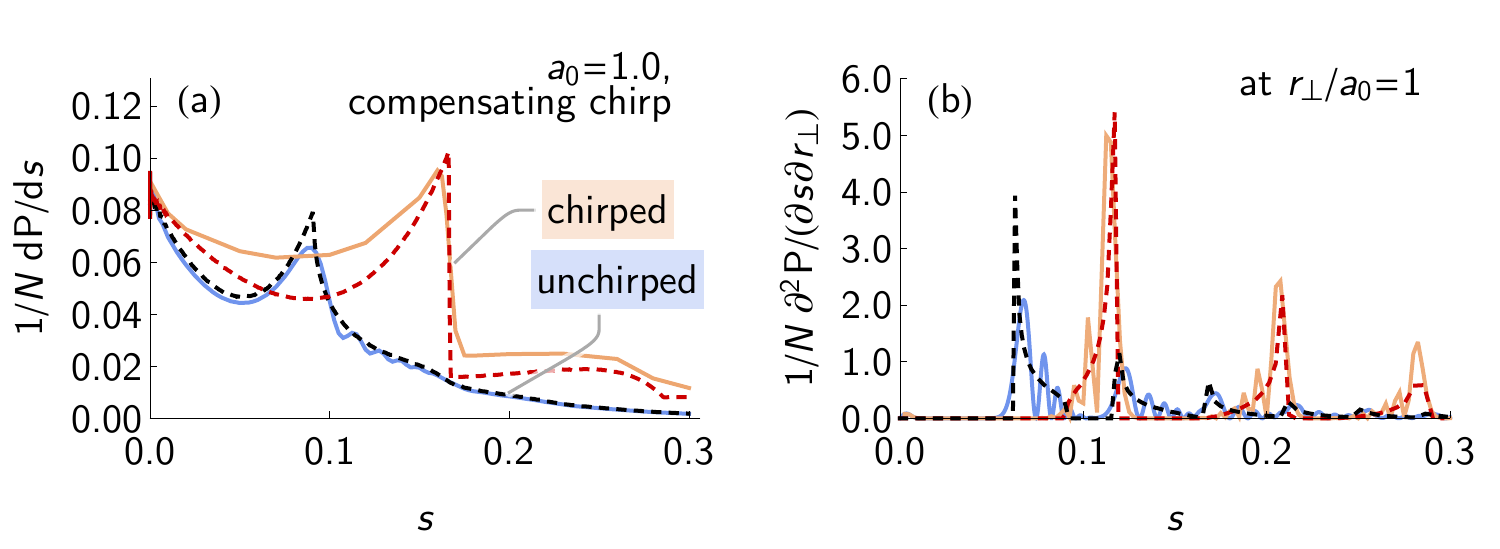}
    \caption{%
        Comparison between simulation (dashed) and QED (solid) results for a pulse with a nonlinear chirp that compensates for the classical redshift (red/orange).
        Here $a_0 = 1$, $N = 16$ and the electron energy parameter $\eta_{0} = 0.1$.
        }
    \label{fig:CompensatingChirp}
    \end{figure*}
    
The second example we present is that of a highly nonlinear chirp, where the instantaneous frequency varies in such a way as to compensate for the classical broadening of the photon spectrum at $a_0 > 1$.
In a pulsed plane wave, the position of the first harmonic edge varies from $s = 2 \eta_{0} /(1 + 2 \eta_{0})$ to $s = 2 \eta_{0} /(1 + a_0^2 + 2 \eta_{0})$ as the cycle-averaged potential $\newa(\phi)$ sweeps up and down.
As such, the on-axis emission is broadband unless the intensity is rather low.
In order to overcome this, and obtain a narrowband source of Compton $\gamma$ rays even when $a_0$ is not small, it has been proposed to chirp the pulse in a particular way~\cite{gsu.prstab.2013,terzic.prl.2014,rykovanov.prab.2016,seipt.prl.2019,tang.arxiv.2021}.
If the instantaneous frequency of the pulse varies as $\omega(\phi) = \omega_0 [1 + \newa^2(\phi)]$, then $s = 2 \eta_0 / (1 + 2 \eta_0)$ for all $\phi$ and the nonlinear redshift is perfectly compensated.
Although there are significant obstacles to achieving this in experiment, it is a useful test case for the simulation method we have introduced.
We therefore consider a pulse with envelope $f(\phi) = \cos^2[\phi/(2N)]$ for $\abs{\phi} < \pi N$ and $b(\phi) = \phi + a_0^2 \int_{0}^\phi f^2(y) \,\rmd y$.
In this case, the chirp may be considered to be slowly varying if $2 a_0^2 / [N (1 + a_0^2)] \ll 1$.
We show results for $a_0 = 1$, $N = 16$ in \cref{fig:CompensatingChirp}.
The lightfront momentum spectrum for theory and simulation both show a shift of the edge of the first harmonic from the nonlinear, to the linear position, as expected for this choice of chirp.
However, this rather extreme choice of chirp leads to a larger discrepancy in the in the height of the spectra: the simulations underestimate the total yield by a small but not insignificant amount.
We have verified that both theory curves tend to the same value in the limit of vanishing $s$, and that the simulation curves do as well: the limiting value, $\lim_{s \to 0}\frac{\rmd\Prob}{\rmd s} \propto a_0^2 \int\! f^2(\phi) \,\rmd\phi$, is sensitive only to the pulse \emph{envelope} (for circular polarization)~\cite{dipiazza.pra.2018,heinzl.pra.2020}.

\section{Focused lasers}

Theoretical calculations of strong-field QED effects in experimentally relevant scenarios must deal with three-dimensional effects:
the nonlinear regime $a_0 \gtrsim 1$ is reached by focusing laser light to a spot of small, even diffraction-limited, size, so the laser pulse will differ significantly from a plane wave;
the electron beam that probes the laser will also have finite size and temporal duration.
Theoretical results build upon analytical solutions of the Dirac equation in a background field and are therefore only available for plane waves, focusing models of very high symmetry~\cite{heinzl.prl.2017,heinzl.jpa.2017}, or under a high-energy approximation $\gamma \gg a_0$~\cite{dipiazza.prl.2014,dipiazza.pra.2017}.
In this section, we discuss the application of simulations, based on LMA emission rates, to model the interaction of electron beams with focused laser pulses.

Within the LMA, the field is treated locally as a monochromatic plane wave.
In order to model a focused laser pulse, we therefore promote the cycle-averaged amplitude $\newa$ and wavevector $\kappa$ to be functions of spatial coordinate as well as phase.
For Gaussian focusing, within the paraxial approximation, we have
    \begin{align}
    \newa &= \frac{a_0 f(\psi)}{\sqrt{1 + \zeta^2}}
        \exp\!\left( -\frac{\rho^2}{1 + \zeta^2}\right),
    &
    \rho^2 &= \frac{x^2 + y^2}{w_0^2},
    &
    \zeta &= \frac{z}{z_R}, \label{eqn:parax1}
    \end{align}
where $w_0$ is the beam waist (the radius at which the intensity falls to $1/e^2$ of its central value), $z_R = \pi w_0^2 / \lambda$ is the Rayleigh range, and the factor $f(\psi)$ is the pulse envelope~\cite{porras.pre.1998}.
The local wavevector $\kappa_\mu = \partial_\mu \psi$, where $\psi = \phi - \rho^2 \zeta / (1 + \zeta^2) + \tan^{-1}\!\zeta$ is the total phase. However, in what follows we neglect the wavefront curvature and Gouy phase so that $\psi = \phi$ and $\kappa$ takes its usual, plane-wave value.
We compare the results so obtained with simulations based on the LCFA, which is a more standard approach~\cite{ridgers.jcp.2014,gonoskov.pre.2015}.
In the LCFA simulations, the laser pulse is defined using the paraxial solution for the fields given in \cite{salamin.apb.2007}: we include terms up to fourth-order in the diffraction angle $\varepsilon = w_0 / z_R$ in the Gaussian beam, which is then multiplied by a temporal envelope $f(\phi)$.
Electron trajectories are determined by solution of the ponderomotive force equation, \cref{eq:PonderomotiveForce}, for the quasimomentum, or the Lorentz force for the kinetic momentum, as appropriate.

\begin{figure*}
    \centering
    \includegraphics[width=0.8\linewidth]{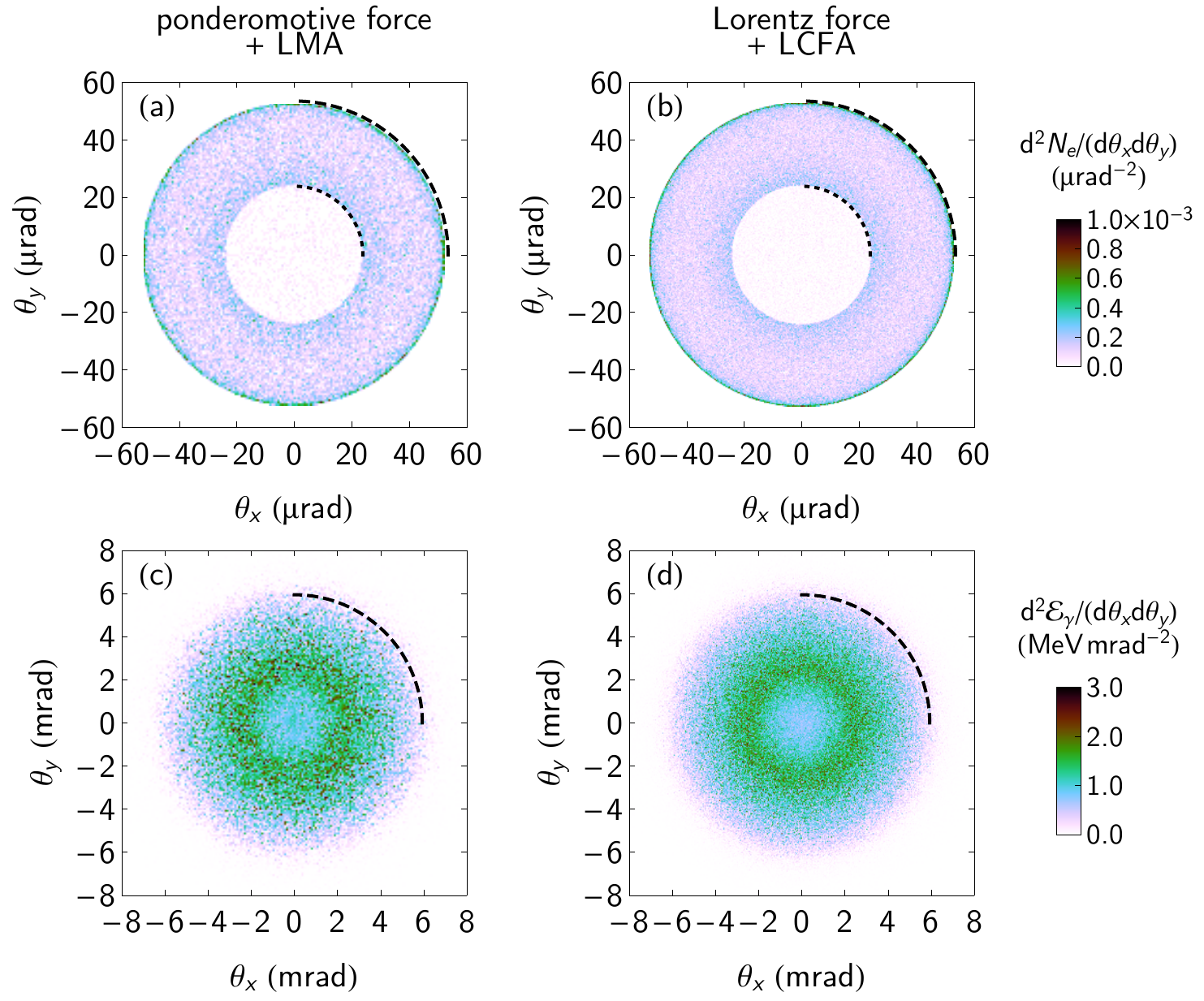}
    \caption{%
        Electron (upper row) and photon (lower row) angular distributions, from LMA- and LCFA-based simulations of an electron beam colliding with a focused laser pulse, with recoil disabled.
        Here the laser pulse has a peak amplitude of $a_0 = 10$, a duration of $30$~fs, and a focal spot size of $w_0 = 2~\micron$.
        The electrons in the beam have energy parameter $\eta_{0} = 0.01$, zero initial divergence, and are distributed uniformly over a disk of radius $r = w_0$.
        Black, dashed lines gives analytical estimates for the scattering angles: see text for details.
        }
    \label{fig:LMAvLCFA}
    \end{figure*}

First, we verify that LMA and LCFA simulations yield consistent results in a regime where they are expected to do so.
We consider a laser pulse that is focused to a spot size $w_0 = 2~\micron$, reaching a peak amplitude of $a_0 = 10$, with Gaussian temporal envelope of (full width at half maximum) duration 30~fs.
The electrons have initial energy parameter $\eta_{0} = 0.01$ (equivalent to $\gamma_0 = 1638$, given a laser wavelength of $0.8~\micron$) and are initially counterpropagating, with zero initial divergence.
Their initial positions are distributed over a disk of radius $r_0 = w_0$, such that they encounter a range of peak intensities.
We have both $a_0 \gg 1$ and $a_0^2 / \eta_{0} \gg 1$, so the LCFA is expected to be a good approximation.
The results presented in \cref{fig:LMAvLCFA} are obtained from simulations of this scenario using the LMA and LCFA, with recoil on photon emission artificially disabled.
This means that the electron trajectory is determined solely by the action of the laser fields, allowing us to confirm the equivalence between the LMA and LCFA at the level of the electron dynamics, illustrated in \cref{fig:Concept}.

\Cref{fig:LMAvLCFA} shows the angular distributions of the electrons and emitted photons, after the collision has taken place.
We see immediately that the LMA and LCFA simulations yield almost identical results.
In order to explain the double ring structure evident in the electron distributions, we derive an approximate, analytical prediction for the expected ponderomotive scattering angle\footnote{Analytical predictions for the scattering angle are also given in~\cite{mackenroth.njp.2019}, but these are derived under the assumptions that the laser transverse intensity profile is flat up to a radius equal to the waist, and that the pulse duration is infinitely long. Neither condition applies here.}.
Consider an electron that is initially counterpropagating, with no initial transverse momentum, at radial distance (impact parameter) $b$ from the laser axis, at ultrarelativistic velocity such that $q^0 \simeq -q^3 \gg q^\perp$.
We approximate $\newa^2 \simeq [a_0 \exp(-r^2/w_0^2) f(\phi)]^2$ and solve the equation of motion, \cref{eq:PonderomotiveForce}, perturbatively in the small parameter $\epsilon \equiv 1/\gamma_0$.
The first-order correction to the perpendicular momentum $q^\perp$ is obtained by substituting into \cref{eq:PonderomotiveForce} $q^0 = m \gamma_0$ and $r = b$, i.e. assuming the electron is undeflected.
The deflection angle follows as $\theta \simeq q^\perp / q^0$:
	\begin{equation}
	\theta_e \simeq
		\frac{a_0^2}{\gamma^2_0}
		\frac{b e^{-2 b^2/ w_0^2}}{z_R}
		\int_{-\infty}^\infty f^2(\phi) \,\rmd\phi.
	\label{eq:DeflectionAngle}
	\end{equation}
The outer ring in \cref{fig:LMAvLCFA}(a) and (b) corresponds to scattering at $b = w_0 / 2$ (shown by the black, dashed line), at which \cref{eq:DeflectionAngle} is maximised, and the inner ring to scattering at $b = w_0$ (shown by the black, dotted line), which is the radius of the electron beam.

As discussed in \cref{sec:Simulations}, and shown in \cref{fig:Concept}, angular structure in the photons emerges differently in the LMA and LCFA simulations.
In the former, it is the emission rate and the conservation of quasimomentum that ensures that photons are most probably emitted at angles $\theta_\gamma \simeq a_0 / \gamma_0$ to the instantaneous quasimomentum.
In the latter, it arises from the instantaneous oscillation in the electron kinetic momentum, which has characteristic angle $\theta_e \simeq a_0 / \gamma_0$, and the fact that the radiation is beamed parallel to this.
The azimuthal symmetry of a circularly polarised laser means that the radiation angular profile is annular in shape: while this is evident in \cref{fig:LMAvLCFA}(c) and (d), the characteristic angle is smaller than the expected value $\theta_\gamma = a_0 / \gamma_0$, which  is shown by the black, dashed line.
This is caused by the fact that the electrons are distributed over a range of impact parameters and therefore encounter lower effective values of $a_0$: $a^\text{eff}_0(b) \simeq a_0 \exp(-b^2/w_0^2)$.

Focal spot averaging not only lowers the yield of photons, as compared to a plane wave with the same peak amplitude, it also reduces the clarity of signatures of strong-field QED effects.
We demonstrate this in particular for the position of the first nonlinear Compton edge, at $a_0 \sim 1$, $\eta_{0} = 0.1$.
This also provides an opportunity to crosscheck our LMA simulation results for focused lasers with theory.
The latter is obtained using \cref{eq:Prob2d}, i.e. under the high-energy approximation that the electron is undeflected during its passage through the laser pulse.
We have already shown that the total deflection angle scales as $(a_0/\gamma_0)^2$, which is indeed very small.
In this case, the laser amplitude is either $a_0 = 0.5$ or $2.5$, its waist is $w_0 = 4~\micron$, and its temporal envelope (electric-field) is $f(\phi) = \cos^2[\phi/(2N)]$ with $N = 16$.
The electrons have energy parameter $\eta_{0} = 0.1$ (equivalent to $\gamma_0 = 1.638 \times 10^{4}$ for a head-on collision with a laser pulse of central wavelength $\lambda = 0.8~\micron$) and are distributed uniformly over a disk of radius $2 w_0$.

    \begin{figure*}
    \centering
    \includegraphics[width=0.8\linewidth]{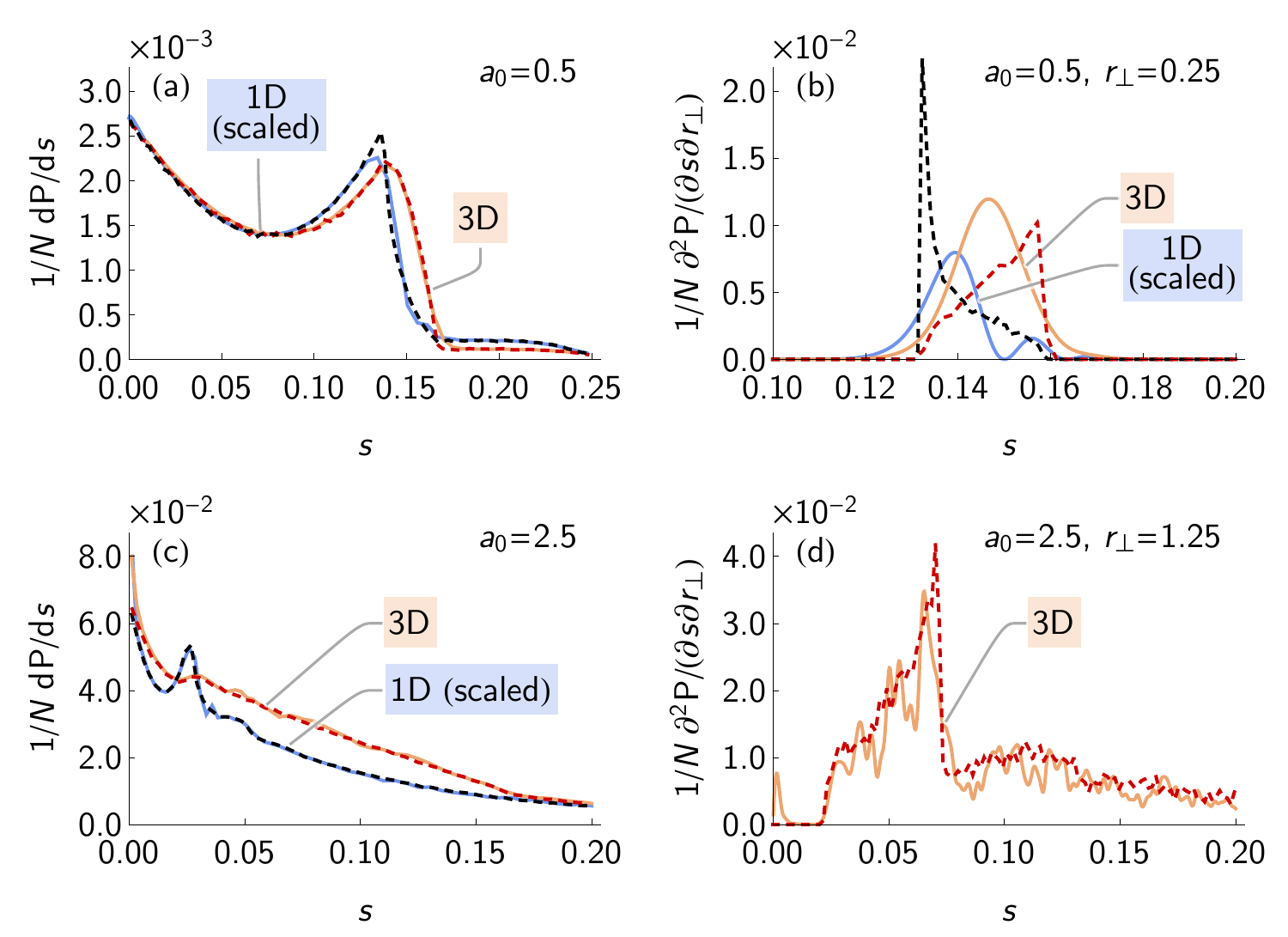}
    \caption{%
        Comparison between simulation (dashed) and theory (solid, coloured) results for a plane wave (blue) and a focused pulse (waist $w_0 = 5\lambda$, orange) with $a_0 = 0.5$ (upper row) and $2.5$ (lower row).
        The pulse duration is $N = 16$ and the electron energy parameter $\eta_{0} = 0.1$.
        In the 3D case, the electrons are initially uniformly distributed over a disk of radius $2 w_0$.
        The 1D results are scaled by a factor $R_\text{3D} = (1 - e^{-8})/8 \simeq 0.125$ (see text for details).
        }
    \label{fig:3D}
    \end{figure*}
    
In \cref{fig:3D}, we compare the theory and simulation results with those obtained for a plane wave with the same peak amplitude.
As the total yield is reduced in the former case, we scale the plane-wave results by a factor $R_\text{3D}$ which approximately accounts for the effect of focal spot averaging.
In the perturbative limit $\newa \ll 1$, the emission rate is proportional to $\newa^2$.
Thus we expect the overall number of photons, in the 3D case, to be reduced by a factor $R_\text{3D} \simeq \left( \int\! \newa^2(b) \frac{\rmd N_e}{\rmd b} \,\rmd b \right) / a_0^2$, where $\frac{\rmd N_e}{\rmd b}$ is the distribution of electron impact parameters $b$, and we may take $\newa(b) = a_0 \exp(-b^2/w_0^2)$ for beam waist $w_0$.
For a beam of electrons which are uniformly distributed over a disk of radius $2 w_0$, we have $R_\text{3D} = (1 - e^{-8})/8 \simeq 0.125$.
The distribution of photon lightfront momentum fraction $s$ is shown in \cref{fig:3D}(a) and (c) for $a_0 = 0.5$ and $2.5$ respectively.
\Cref{fig:3D}(b) and (d) show lineouts through the double-differential spectrum at fixed $r_\perp = a_0 / 2$.
The agreement between theory and simulation is reasonably good.
The detailed structure in the lineouts is not resolved, because the LMA misses interference effects at the scale of the pulse envelope. This is more evident in \cref{fig:3D}(b) than (d), i.e. at lower $a_0$, for the following reason.
In the LMA, the only contribution to the bandwidth of an individual harmonic is the variation in the mass shift over the pulse duration: at fixed $r_\perp$, this width is $\Delta s / s = a_0^2 / (1 + 2 n \eta)$.
There is an additional contribution from the non-zero bandwidth of the pulse, which is given approximately by $\Delta s / s \simeq 0.187 / (\omega_0 \tau)$, where $\tau$ is the FWHM duration of the pulse intensity profile: for the cosine-squared envelope under consideration here, $\omega_0 \tau \simeq 0.364 N$ and $\Delta s / s \simeq 1/(2N)$.
At sufficiently small $a_0$, it is the latter contribution, from the laser pulse bandwidth, that dominates.
Note that in a focussed pulse, the effective amplitude at finite impact parameter $a_\text{rms}(b) < a_0$ and so such effects are magnified.
Integrating over a finite range of $r_\perp$ partially mitigates this, which is why the single-differential spectra are in much better agreement.

The difference between the 1D and 3D cases, evident in the theory, is captured very well by the simulations.
We see that the first nonlinear edge is smeared out by focal spot averaging, particularly for $a_0 = 2.5$.
This is because the position of the edge differs for electrons at different impact parameters, as increasing $b$ means reducing the effective $a_0$.
We have repeated the comparison between LMA-based simulations and QED for more tightly focussed laser pulses, reducing the waist $w_0$ to $3\lambda$ and $1\lambda$, while holding the peak $a_0$ and the electron-beam--laser overlap fixed.
The detailed results are shown in the Supplementary Material: we find that the spectra are barely affected by the reduction and the agreement between simulations and theory is consistently good.
This supports our expectations that LMA-based simulations are accurate even for focussed laser pulses.

\section{Discussion}
\label{sec:Discussion}

    \begin{table}
    \centering
    \begin{ruledtabular}
    \begin{tabular}{ccc}
        Predicted quantity & LCFA is accurate when & LMA is accurate when \\
        Yield, $N_\gamma$ & \makecell{$a_0^2/\eta \gg 1$, $a_0 \gg 1$, $\omega_0 \tau \gg 1$}\footnote{The last condition expresses that the photon formation length should be smaller than both the wavelength $\lambda$ and the pulse length $\tau$. It follows from the standard LCFA applicability conditions $[(a_0^2/\eta) s/(1-s)]^{1/3} \gg 1$ and $a_0 \gg 1$, when one chooses $s = s^\star_{0}$ and $\eta \ll N a_0^2$, corresponding to the mid-IR peak.} & \makecell{$\omega_0 \tau \gg 1$} \\
        Spectrum, $\displaystyle\frac{\rmd N_\gamma}{\rmd s}$ & \makecell{$\displaystyle s \gtrsim \frac{2\eta}{1+a_0^2+2\eta}$} & \makecell{$s \gtrsim s^\ast_0$, $s \ll s^\ast_0$ \\ where $s^\ast_0 = \frac{\eta / N}{1+a_0^2+ \eta / N}$}
    \end{tabular}
    \end{ruledtabular}
    \caption{Conditions on the pulse amplitude $a_0$, phase duration $\omega_0 \tau$ and electron energy parameter $\eta = \kappa.p / m^2$ for simulated photon yields and spectra to be accurate under the LMA and LCFA.}
    \label{tbl:Validity}
    \end{table}

The focus of this paper has been incorporating the LMA into numerical simulation and providing the first benchmarks for a range of parameters and observables with direct calculation from QED.
As part of this work, we compare the LMA to the LCFA, which is the standard scheme for including QED effects in the modelling of intense laser interactions.
The power of the LCFA is, in part, due to its versatility.
It can be used when the strong electromagnetic field is not known \emph{a priori}, which is a particular advantage when dealing with a laser-plasma interaction.
However, in situations where the shape of the intense laser pulse is well-known, and unchanged in the interaction, the LMA can be used to attain a higher precision than the LCFA.
The demand for the precision is acute if the field strength and particle energies in question are outside the region of validity of the LCFA, as is the case in some upcoming high-energy experiments~\cite{luxe,meuren.exhilp.2019}.

Using a plane-wave pulse with phase duration $\omega_0 \tau = 2 \pi N$, we can give some indication for parameter regimes where the approximations can be used. We designate the region of accuracy as being when particle and field parameters are far away from values where it is known that the approximation is in doubt.
We summarize our findings in \cref{tbl:Validity} for the photon yield and spectrum separately, as the conditions depend on the quantity that is to be measured.
Note that, for these two quantities, the `penalty' from violating the validity conditions is not equal in each case: if the LCFA is used to calculate the yield outside of the range given by $a_{0}$ and $\eta$, the prediction can be wrong by orders of magnitude (as demonstrated by \cref{fig:Probability}).
The requirement that $\omega_0 \tau \gg 1$ for the LMA to be accurate is softer.
\Cref{fig:Probability} shows that the simulations are accurate to within a few per cent even if $N = 4$, which is already much shorter than most typical laser pulses (where the equivalent $N \gtrsim 10$): furthermore we expect the accuracy to improve with increasing $N$.
Similarly, the error made by the LCFA in the photon spectrum becomes arbitrarily large as $s \to 0$, even if $a_0 \gg 1$, whereas the LMA is guaranteed to obtain the correct limiting value for all $a_0$.
The LMA result is inaccurate only in a small region around $s \simeq s_0^\ast = \frac{\eta / N}{1+a_0^2+ \eta / N}$~\cite{king.prd.2021}, which is in any case missed by the LCFA: radiation here arises from the slow ponderomotive scattering of the electron.

If the plane wave contains a chirp, then the condition that the LMA is still a good approximation is found to be $|\omega'/\omega| \ll 1$, which is a `slowly varying' condition of the same nature as $1/N \ll 1$.
In practice, one is often limited to such chirps by virtue of the available bandwidth~\cite{silvestro.jqe.1984}.
For focused laser pulses, we may expect that the applicability regime apparent when a chirp was introduced, should be adaptable for the change in wave vector $\vec{k}$ obeying $|\vec{k}'|/|\vec{k}| \ll 1$.
This reduces to a condition on the focusing, which may be expressed through the diffraction angle $\varepsilon = \lambda / (\pi w_0)$ as $\varepsilon \ll 1$. 
(A similar condition applies to the validity of the ponderomotive-force approach to the particle's classical dynamics~\cite{quesnel.pre.1998}, which confirm for $\varepsilon \simeq 0.12$ in \cref{fig:3D}.)
Using the high-energy approximation with the focused background in \cref{eqn:parax1}, we note $z/z_R = -\phi/(2 \omega z_R)$, and the relevant change in wave-vector on the focal axis is $k'/k \sim 1/(\omega z_R) = \varepsilon^{2}/2 \ll 1$.
We have cross-checked our simulation results against a theoretical calculation which uses a WKB approximation, suitable for high-energy electrons.
This integrates the plane-wave probability over a transverse distribution of intensity and particle flux and therefore works in a similar way to our simulations.
We find good agreement even for diffraction-limited focal spots (see Supplementary Material), when $|\vec{k}'|/|\vec{k}| \ll 1$ still holds.
To benchmark the LMA fully would require solutions to the Dirac equation in backgrounds beyond a plane wave, which is still an active area of research \cite{Bagrov:1990xp,PhysRevD.94.065039,PhysRevLett.118.113202}.

\section{Summary}

Motivated by the imminent need for precision simulations of strong-field QED processes in the transition regime $a_0 \sim 1$, we have presented here a novel simulation framework which incorporates quantum effects via probability rates calculated within the locally monochromatic approximation (LMA)~\cite{heinzl.pra.2020}.
From the theory perspective, the formalisation of the LMA from the plane-wave model has been extended to include chirped pulses, under a ``slowly varying chirp'' approximation.
We have also adapted the LMA to model focused laser backgrounds, under the approximation that the incident electron has a relativistic $\gamma$ factor satisfying $\gamma \gg a_{0}$.

The emission rates so derived are embedded within a classical  simulation framework that assumes a definite particle trajectory.
In contrast to simulations based on the locally constant field approximation (LCFA), the electron quasimomentum (the cycle-averaged kinetic momentum) plays the essential role here, appearing in the classical equations of motion and the conservation of momentum.
The fast oscillation of the particle momentum, at the timescale of the laser frequency, is nevertheless included, but at the level of the emission rates.
This simulation framework therefore has conceptual similarities to the ``envelope solvers'' used to model laser-wakefield acceleration~\cite{mora.pop.1997,cowan.cpc.2011,benedetti.ppcf.2017}.

In benchmarking the simulations against QED results, we have found excellent agreement for a variety of background field configurations.
Furthermore, we obtain significant reductions in the relative error when compared to the use of the LCFA in the transition regime.
While we have focused, in this work, on the specific example of nonlinear Compton scattering in a circularly polarised background, our results can be extended to other processes, such as electron-positron pair creation~\cite{ritus.jslr.1985,heinzl.pra.2020}, and to include spin- and polarisation-dependence~\cite{king.pra.2013,delsorbo.pra.2017,wistisen.prd.2019,king.pra.2020b,seipt.arxiv.2020}.

\begin{acknowledgments}
We would like to thank members of the LUXE collaboration for helpful discussions during preparation of this work.
We acknowledge funding from the Engineering and Physical Sciences Research Council (grant EP/S010319/1, B.K., A.J.M.).
Simulations were performed on resources provided by the Swedish National Infrastructure for Computing at the High Performance Computing Centre North.
\end{acknowledgments}

\section*{Data availability}
The source code for the simulation program described is available at Ref.~\cite{ptarmigan}.
Version 0.6.0, which is used in this work, the input configurations necessary to reproduce the simulation results, and the analytical results, are archived at Ref.~\cite{dataset}.

\appendix

\section{Locally monochromatic approximation for general chirped plane-wave pulses \label{app:LMA}}

In \cite{heinzl.pra.2020}, the LMA was derived from plane-wave QED for a simple plane-wave pulse. A plane wave is a highly idealised model of a laser field, which does not take into account some of the important characteristics of pulses in a real experiment.
Here we extend the LMA to the case of a plane-wave pulse which includes an arbitrary chirp. We begin with a general overview of the LMA for a plane-wave field with a general chirp term. 

For concreteness, we use a circularly polarised pulse with an arbitrary chirp, where the dimensionless gauge potential $a_{\mu}(\varphi) = e A_{\mu}(\varphi)/m$ is
    \begin{equation}
    a_\mu(\phi) =
        a_0
        f\!\left(\frac{\varphi}{\Phi}\right)
        \left[\eps_\mu \cos b(\varphi) + \beta_\mu \sin b(\varphi)\right],
    \label{eqn:Circular}
    \end{equation}
and the phase is $\varphi = \kappa \cdot x$.
In the derivation of the LMA, it is more natural to work with functions of the phase variable $\varphi$, than the proper time $\tau$, which is used in the main text, and so in what follows we work with $\varphi$.
The discussion here can be generalised to linearly or elliptically polarised backgrounds (see \cite{heinzl.pra.2020} for more details on the subtleties involved in the LMA for a linear, unchirped, plane-wave pulse).

We follow the standard approach of defining the scattering amplitude for our process in terms
of the Volkov wavefunctions for the background dressed fermions of mass
$m$ and 4-momentum $p_{\mu}$,
\cite{Volkov:1935zz},
\begin{align}\label{def:Volkov}
    \Psi_{p,r}(x)
    =
    \left(
        1
        +
        \frac{m \slashed{\kappa} \slashed{\gaugea}(\varphi)}{2 \kappa \cdot p}
    \right)
    u_{p,r}
    e^{- i S_p(x)} 
    \;,
\end{align}
where $u_{p}$ are constant spinors. The Volkov phase term is given by,
\begin{align}\label{def:ClassAct}
    S_p(x)
    = 
    p \cdot x
    +
    \int^{\varphi}_{-\infty}
    \! \rmd y\,
    \frac{2 m p \cdot \gaugea(y) - m^2 \gaugea^2(y)}{2 \kappa \cdot p},
\end{align}         
which is just the classical action for an electron in a plane-wave
background field. The nontrivial dependence of the Volkov wavefunctions
on the phase $\varphi$ means that overall momentum conservation for an
arbitrary scattering amplitude $\tsf{S}$ in the plane-wave background
field only holds for three of the four directions, $\{-, \perp\}$. As
such, the scattering amplitude takes the form,
\begin{align}\label{def:Amplitude}
    \tsf{S}
    =
    (2\pi)^3
    \delta^3_{-,\perp}(p_{\text{in}} - p_{\text{out}})
    \mcM
    \;,
\end{align}
where $\delta^{3}_{-,\perp}(p) =
\delta(p_{\lcm})\delta(p_{1})\delta(p_{2})$, and $\mcM$ is the invariant
amplitude.

Closed form solutions to \cref{def:ClassAct} are {not always available. A simple example is the} infinite monochromatic plane wave, which is the $f(\varphi/\Phi) \to 1$, $b(\varphi) \to \varphi$ limit of the background field \cref{eqn:Circular}.
However, one can separate the fast and slow dynamics of the background field in such a way that the field dependent terms in the exponent can by integrated by parts, and simplified by neglecting derivative corrections. This technique is known as the slowly varying envelope approximation~\cite{heinzl.pra.2020,Narozhnyi:1996qf,McDonald:1997,seipt.pra.2011,seipt.jpp.2016}.

The slowly varying envelope approximation for an arbitrarily chirped
plane-wave field was derived in \cite{seipt.pra.2015}, and we follow this
approach here. For the circularly polarised background
\cref{eqn:Circular}, the terms which are quadratic in the field depend
only on the slowly varying envelope, $\gaugea^{2}(\varphi) = -a_{0}^{2} f^{2}(\varphi/\Phi)$, while the terms linear in the field
contain both slow (through $f$) and fast (through $b$) timescales. This
gives integrals of the form,
\begin{align}\label{eqn:Integrals}
    \mcI
    =
    \int^{\varphi}_{-\infty}
    \! \rmd y \,
    f \left( \frac{y}{\Phi} \right)
    \left[
        \cos b(y),
        \sin b(y)
    \right]
    .
\end{align}
To deal with these integrals, we first transform the trigonometric functions of $f(y)$ to pull out a factor depending on the inverse of $\omega(y) = b^{\prime}(y)$, where a prime denotes {a derivative of the argument}:
\begin{align}\label{eqn:IntegralsTransform}
    \mcI
    =
    \int^{\varphi}_{-\infty}
    \! \rmd y \,
    \frac{f(y/\Phi)}{\omega(y)}
    \frac{\rmd}{\rmd y}
    \left[
        \sin b(y),
        -\cos b(y)
    \right].
\end{align} 
The function $\omega(y)$ is taken to define a \emph{local frequency
scale}. Each term can then be readily integrated by parts, giving
two contributions: a boundary term and a term proportional to
\begin{align}\label{eqn:Derivative}
    \frac{\rmd}{\rmd y}
    \frac{f(y/\Phi)}{\omega(y)}
    =
    \frac{1}{\Phi}
    \frac{f^{\prime}(y/\Phi)}{\omega(y)}
    -
    \frac{f(y/\Phi)}{\omega^{2}(y)}
    \omega^{\prime}(y).
\end{align}
Provided this is a small correction, which is valid for sufficiently
long pulses, $\Phi \gg 1$ and when the derivative of the chirp function
satisfies $\omega^{\prime}(y) \ll \omega(y)$, we can neglect these slowly varying
terms, and approximate the integrals by,
\begin{align}\label{eqn:IntegralsApprox}
    \mcI
    \simeq
    \frac{f(\varphi/\Phi)}{\omega(\phi)}
    \left[
        \sin b(\varphi),
        -\cos b(\varphi)
    \right].
\end{align}
Applying these approximations to the classical action $S_{p}$ in
\cref{def:ClassAct} gives,
\begin{align}\label{eqn:ClassActApprox}
    S_{p}(x)
    =
    G(\varphi)
    -
    z(\varphi)
    \sin\!\left[ b(\varphi) - \vartheta \right].
\end{align}
The function $G(\varphi)$ contains only slowly varying terms, or terms
linear in $\varphi$. The function $z(\varphi)$ depends on the phase
only through the slowly varying envelope $f(\varphi/\Phi)$ and local
frequency $\omega(\varphi)$, and the angle $\vartheta$ is \emph{independent} of the phase.

The exponential of the trigonometric function in
\cref{eqn:ClassActApprox} can be expanded into an infinite sum of
Bessel functions using the Jacob-Anger expansion,
\begin{align}\label{eqn:Bessel}
    e^{- i z \sin(b - \vartheta)}
    =
    \sum_{n = - \infty}^{\infty}
    e^{- i n (b - \vartheta)}
    J_{n}(z)
    \;.
\end{align}
For the case of a one vertex process, such as nonlinear Compton scattering or Breit-Wheeler pair production, once the oscillating phase term has been expanded by \cref{eqn:Bessel}, the invariant amplitude, $\mcM$, in \cref{def:Amplitude}, takes on the form,
\begin{align}\label{eqn:InvAmplitude}
    \mcM
    =
    \int \! \rmd \varphi
    \sum_{n = - \infty}^{\infty}\!
    {\mcM_{n}'(\varphi)}
    \;.
\end{align}

The probability, $\Prob$, is then found in the usual way by squaring the
scattering amplitude \cref{def:Amplitude} and integrating over the
Lorentz invariant phase space for the particular process,
$\rmd\Omega_{\text{LIPS}}$,
\begin{align}\label{eqn:GeneralProb}
    \Prob
    \propto
    \iint \rmd \varphi \, \rmd\varphi^{\prime}
    \sum_{n,n^{\prime} = - \infty}^{\infty}
    \int \! \rmd\Omega_{\text{LIPS}}
    \,
    {\mcM_{n}^{'\,\dagger}(\varphi)}
    {\mcM^{'}_{n^{\prime}}}(\varphi^{\prime}).
\end{align}
There are now two phase integrals, and what distinguishes the LMA from
the slowly varying approximation (which is all we have applied
so far) is performing a local expansion in the phase variables. To
achieve this we introduce the sum and difference variables,
\begin{align}\label{eqn:SumDiff}
    \phi 
    =
    \frac{1}{2} (\varphi + \varphi^{\prime})
    \;,
    \qquad \qquad
    \theta
    =
    \varphi - \varphi^{\prime}
    \;,
\end{align}
and then take the small phase difference approximation $\theta \ll 1$ to expand the probability in a Taylor series in $\theta$, retaining only the leading-order, $O(\theta)$, contributions.

The $\theta$-integral can be performed analytically, leaving the
probability in the form,
\begin{align}\label{eqn:ProbabilityStructure}
    \Prob^\text{LMA}
    =
    \int \!\rmd\phi \,
    \tsf{R}^\text{LMA}(\phi).
\end{align}
The function, $\tsf{R}^\text{LMA}(\phi)$, contains summations over the Bessel harmonics and integrations over the final states, but crucially only depends on \emph{one} phase variable. 
This allows us to interpret $\tsf{R}(\phi)$ as a \emph{local rate} which can be used in simulations. (In the main paper, we instead use a rate $W^{\trm{LMA}}$ defined as a probability per unit proper time.) 
To make this discussion more explicit, we consider the process of nonlinear Compton scattering.

\subsection{Nonlinear Compton scattering in a chirped plane-wave pulse \label{sec:NLC}}

Consider an electron with an initial momentum $p_{\mu}$ interacting with a plane-wave electromagnetic field to produce a photon of momentum $k_{\mu}$ and polarisation $\epsilon_{k,l}^{*}$. The scattering amplitude, in terms of the Volkov wave functions \cref{def:Volkov}, is given by,
\begin{align}\label{eqn:NLCAmplitude}
    \tsf{S}_{r',r;l}    
    =
    -
    i
    e
    \int \! \rmd^{4}x \,
    \bar{\Psi}_{p^{\prime},r^{\prime}}(x)
    \slashed{\epsilon}^{*}_{k,l}
    e^{i k \cdot x}
    \Psi_{p,r}(x).
\end{align}
Here we use the Dirac slash notation, $\slashed{\epsilon} =
\gamma^{\mu}\epsilon_{\mu}$, where $\gamma^{\mu}$ are the Dirac gamma matrices.
The momentum $p_{\mu}^{\prime}$ is the momentum of the outgoing electron. 

Performing all of the trivial integrations to express the scattering amplitude
in the form \cref{def:Amplitude}, the invariant amplitude is found to be,
\begin{align}\label{eqn:NLCInvariant}
    \mcM
    =
    -
    i
    e
    \int \!\rmd\varphi \,
    \mcS(\varphi)
    \exp\!\left[
        i 
        \int^{\varphi}_{-\infty}
        \! \rmd y \,
        \frac{k \cdot \pi(y)}{\kappa \cdot (p - k)}
    \right].
\end{align}
where the spin dependent structure is given by,
\begin{align}\label{eqn:Spin}
    \mcS(\varphi)
    =
    \bar{u}_{p^{\prime},r^{\prime}}
    \left[
        1
        +
        \frac{m \slashed{\gaugea}(\varphi) \slashed{\kappa}}{2 \kappa \cdot p^{\prime}}
    \right]
    \slashed{\epsilon}_{k,l}^{*}
    \left[
        1
        +
        \frac{m \slashed{\kappa} \slashed{\gaugea}(\varphi)}{2 \kappa \cdot p}
    \right]
    u_{p,r}.
\end{align}
and the classical action in the exponent is expressed in terms of the kinetic,
or local, momentum of the incoming electron,
\begin{align}\label{def:Kinetic}
    \pi^\mu(\varphi)
    =
    p^\mu
    -
    m \gaugea^\mu(\varphi)
    +
    \frac{2 m p \cdot \gaugea(\varphi) - m^2 \gaugea^2(\varphi)}{2 \kappa \cdot p}
    \kappa^\mu
    \;.
\end{align}

After applying the slowly varying approximation, as detailed above, to
the classical action in the exponent, the invariant amplitude
\cref{eqn:NLCInvariant} can be expressed as
\begin{align}\label{eqn:NLCSlowInv}
    \mcM
    =
    -
    i
    e
    \int \!\rmd\varphi \,
    \mcS(\varphi)
    e^{
        i G(\varphi) - i z(\varphi) \sin[b(\varphi) - \vartheta]
    }.
\end{align}
The function $G(\varphi)$ has the explicit form,
\begin{align}\label{eqn:NLCG}
    G(\varphi)
    =
    \frac{1}{2 s \kappa \cdot p (1 - s)} 
    \int^{\varphi}_{-\infty}
    \! \rmd y
    \left\{
        |\vec{k}_{\perp} - s \vec{p}_{\perp}|^{2}
        +
        s^{2} m^{2}
        \left[
            1
            +
            a_{0}^{2}
            f^{2}\!\left( \frac{y}{\Phi} \right)
        \right]
    \right\},
\end{align}
where we have defined the lightfront momentum fraction $s = \kappa \cdot k/\kappa \cdot
p$. As stated above, this only has dependence on the phase through either linear
or slowly varying terms.

The term $z(\varphi)$ is
\begin{align}\label{eqn:NLCz}
    z(\varphi)
    =
    \frac{m a_{0}}{\kappa \cdot p (1 - s)}
    \frac{|f(\varphi/\Phi)|}{|\omega(\varphi)|}
    \sqrt{
        \big|\vec{k}_{\perp} - s\vec{p}_{\perp}\big|^{2}
    }
    \;,
\end{align}
and so the only dependence on the phase comes through the ratio of the
slowly varying pulse envelope and the local frequency. The angle $\vartheta$ is
defined through the relationship,
\begin{align}\label{eqn:vartheta}
    \vartheta
    =
    \arctan\!\left[
        \frac{
            (k - sp) \cdot \varepsilon
        }{(k - sp) \cdot \beta}
    \right],
\end{align}
and so can be interpreted as the angle between the components of the 4-vector
$k_{\mu} - s p_{\mu}$ projected onto the directions of background field
polarisation.

We skip now to the explicit form of the probability. 
Expanding into Bessel harmonics according to~\cref{eqn:Bessel}, the probability~\cref{eqn:GeneralProb} becomes
\begin{align}\label{eqn:NLCProbBessel}
    \Prob^\text{LMA}
    =
    &
    -
    \frac{\alpha m^{2}}{4 \pi^{2} (\kappa \cdot p)^{2}} 
    \iint \!\rmd \varphi \,\rmd\varphi^{\prime}
    \!\sum\limits_{n, n^{\prime} = -\infty}^{\infty}
    \int \!\frac{\rmd s}{s (1 - s)} 
    \int \!\rmd^{2} \vec{k}_{\perp}
    e^{
        i G(\varphi) - i G(\varphi^{\prime})
        - i n b(\varphi) + i n^{\prime} b(\varphi^{\prime}) 
        + i (n - n^{\prime}) \vartheta
    }
    \nonumber\\
    &
    \times
    \left(
        \left\{
            1
            +
            \frac{a_0^2}{2}
            \left[
                1 + \frac{s^{2}}{2 (1 - s)}     
            \right]
            \left[
                f^{2}\Big(\frac{\varphi}{\Phi}\Big) 
                + 
                f^{2}\Big(\frac{\varphi^{\prime}}{\Phi}\Big) 
            \right]
        \right\}
        J_{n}(z(\varphi))
        J_{n^{\prime}}(z(\varphi^{\prime}))
    \right.
        \nonumber\\
    &\left.
    \vphantom{a} -
    \frac{a_0^2}{2}
    \left[
        1 + \frac{s^{2}}{2 (1 - s)}     
    \right]
        f\Big(\frac{\varphi}{\Phi}\Big) 
        f\Big(\frac{\varphi^{\prime}}{\Phi}\Big)
        \left[
            J_{n + 1}(z(\varphi))
            J_{n^{\prime} + 1}(z(\varphi^{\prime}))
            +
            J_{n - 1}(z(\varphi))
            J_{n^{\prime} - 1}(z(\varphi^{\prime}))
        \right]
\right)
\;.
\end{align}

The probability in this form contains two infinite sums over the Bessel
harmonics and integrals over the outgoing photon momentum. Note the
exponential dependence on the chirp function, $b(\varphi)$, and the angle
$\vartheta$. If we consider the definitions
\cref{eqn:NLCG}--\cref{eqn:vartheta}, we notice that the only dependence on
the transverse photon momentum is through the combination $\vec{r}_{\perp} = \vec{k}_{\perp}/(s m) - \vec{p}_{\perp}/m$. We can then shift the integration variables
in \cref{eqn:NLCProbBessel}, and using \cref{eqn:vartheta} express the
integration measure in polar coordinates,
\begin{align}\label{eqn:IntPolar}
    \int \rmd^{2} \vec{k}_{\perp}
    \to
    s^{2} m^2
    \int \rmd^{2} \vec{r}_{\perp}
    =
    \frac{s^2 m^2}{2}
    \int_{0}^{2\pi} \rmd \vartheta
    \int \rmd |\vec{r}_{\perp}|^{2}
    \;.
\end{align}
The only dependence of the probability on the angle $\vartheta$ is then through
the exponential factor $\exp(+i(n - n^{\prime})\vartheta)$. The integration over
the angle $\vartheta$ sets $n = n^{\prime}$. This allows the probability to be well approximated by, 
\begin{align}\label{eqn:NLCProbBesselInt}
    \Prob
    \simeq
    &
    -
    \frac{\alpha m^{4}}{4 \pi (\kappa \cdot p)^{2}} 
    \iint \! \rmd \varphi \, \rmd \varphi^{\prime}
    \!\sum\limits_{n = -\infty}^{\infty}
    \int \!\rmd s \frac{s}{(1 - s)} 
    \int \!\rmd |\vec{r}_{\perp}|^{2}
    e^{
        + i G(\varphi) - i G(\varphi^{\prime})
    }
    e^{- i n (b(\varphi) - b(\varphi^{\prime}))}
    \nonumber\\
    &
    \times
    \left(
        \left\{
            1
            +
            \frac{a_0^2}{2}
            \left[
                1 + \frac{s^{2}}{2 (1 - s)}     
            \right]
            \left[
                f^{2}\Big(\frac{\varphi}{\Phi}\Big) 
                + 
                f^{2}\Big(\frac{\varphi^{\prime}}{\Phi}\Big) 
            \right]
        \right\}
        J_{n}(z(\varphi))
        J_{n}(z(\varphi^{\prime}))
    \right.
        \nonumber\\
    &
    \left.
    \vphantom{a} -
    \frac{a_0^2}{2}
    \left[
        1 + \frac{s^{2}}{2 (1 - s)}     
    \right]
        f\Big(\frac{\varphi}{\Phi}\Big) 
        f\Big(\frac{\varphi^{\prime}}{\Phi}\Big)
        \left[
            J_{n + 1}(z(\varphi))
            J_{n + 1}(z(\varphi^{\prime}))
            +
            J_{n - 1}(z(\varphi))
            J_{n - 1}(z(\varphi^{\prime}))
        \right]
    \right).
\end{align}

Following through with the local expansion, using \cref{eqn:SumDiff} and
$\theta \ll 1$, the integral over $\rmd\theta$ can be performed, which gives a
$\delta$-function:
\begin{align}\label{eqn:NLCProbDelta}
    \Prob
    \simeq
    &
    -
    \frac{\alpha}{\eta_{0}}
    \int \!\rmd \phi
    \sum\limits_{n = 1}^{\infty}
    \int \!\rmd s
    \int \! \rmd |\vec{r}_{\perp}|^{2}
    \delta \!\left[
        |\vec{r}_{\perp}|^{2}
        +
            1
            +
            a_{0}^{2}
            f^{2}\Big(\frac{\phi}{\Phi}\Big) 
        -
        \frac{2 \eta_{0} n \omega(\phi)(1 - s)}{s}
    \right]
    \nonumber\\
    &
    \times
    \left\{
        J_{n}^{2}(z(\phi))
        +
        \frac{a_0^2}{2}
        \left[
            1 + \frac{s^{2}}{2 (1 - s)}     
        \right]
        f^{2}\Big(\frac{\phi}{\Phi}\Big) 
        \Big[
            2
            J_{n}^{2}(z(\phi))
            -
            J_{n + 1}^{2}(z(\phi))
            -
            J_{n - 1}^{2}(z(\phi))
        \Big]
    \right\},
\end{align}
where we have defined $\eta_{0} = \kappa \cdot p / m^{2}$. 
The probability only has support when the argument of the $\delta$-function satisfies:
\begin{align}\label{eqn:DeltaSupport}
    |\vec{r}_{\perp}|^{2}
    +
        1
        +
        a_{0}^{2}
        f^{2}\Big(\frac{\phi}{\Phi}\Big) 
    -
    \frac{2 \eta_{0} n \omega(\phi)(1 - s)}{s}
    =
    0
    \;,
\end{align}
which (upon adapting the notation) is found to be \emph{exactly} the stationary
phase condition which is evaluated in \cite{seipt.pra.2015} (see eq. (25) of \cite{seipt.pra.2015}).
In that work, the stationary phase approximation is carried out at the level of
the amplitude for nonlinear Compton scattering in the slowly varying envelope
approximation. Here we have shown that the exact same kinematic relationship
reappears at the probability level after the explicit application of a
\emph{local expansion}. 

The integral over the remaining perpendicular momentum dependence can be
trivially carried out using the $\delta$-function in \cref{eqn:NLCProbDelta},
which gives the relatively concise expression (suppressing explicit dependence on $\phi$)
\begin{align}\label{eqn:NLCProbFinal}
    \Prob
    \simeq
    &
    -
    \frac{\alpha}{\eta_{0}}
    \int \!\rmd \phi
    \sum\limits_{n = 1}^{\infty}
    \int_{0}^{s_{n,\ast}(\phi)} \!\rmd s
    \left\{
        J_{n}^{2}(z_{n})
        +
        \frac{\newa^2}{2}
        \left[
            1 + \frac{s^{2}}{2 (1 - s)}     
        \right]
        \Big[
            2
            J_{n}^{2}(z_{n})
            -
            J_{n + 1}^{2}(z_{n})
            -
            J_{n - 1}^{2}(z_{n})
        \Big]
    \right\},
\end{align}
where the argument of the Bessel functions is now
\begin{align}\label{eqn:NLCzharm}
    z_{n}(\phi)
    =
    \frac{2 n \newa}{\sqrt{1 + \newa^2}}
    \sqrt{
        \frac{1}{\omega(\phi)}
        \frac{1}{s_{n}(\phi)}
        \frac{s}{1 - s}
        \left[
            1
            -
            \frac{1}{\omega(\phi)}
            \frac{1}{s_{n}(\phi)}
            \frac{s}{1 - s}
        \right]
    },
\end{align}
and we have defined the cycle-averaged potential $\newa = a_{0} f(\phi/\Phi)$ and the upper bound on the integration over $s$ is
\begin{align}\label{eqn:sn}
    s_{n,\ast}(\phi)
    =&
    \frac{s_{n}(\phi) \omega(\phi)}{1 + s_{n}(\phi) \omega(\phi)}
    ,
    \quad&
    s_{n}(\phi)
    =&
    \frac{2 n \eta_{0}}{1 + \newa^{2}}.
\end{align}
Thus, when compared with the expressions found for the LMA in a non-chirped
pulse~\cite{heinzl.pra.2020}, the chirp function, $b(\phi)$, contributes an effective \emph{rescaling} of the lightfront energy parameter, $\eta_{0} \to \eta_{0} \omega(\phi)$, inside the argument of the Bessel functions.
In \cref{eqn:sndef} we have redefined $s_n$ and $s_{n,\ast}$ by absorbing the local frequency, $\omega$ (where $\omega=\omega(\phi)$), into the definition of the local energy parameter, $\eta = \eta_{0} \omega$ (where $\eta=\eta(\phi)$).

\bibliography{references}

\end{document}